\documentclass[journal=jacsat,manuscript=article]{achemso}

\usepackage[version=3]{mhchem} 

\author{Yuling Zhang}
\affiliation{Shanghai Institute of Applied Mathematics and Mechanics, School of Mechanics and Engineering Science, Shanghai Frontier Science Center of Mechanoinformatics, Shanghai Key Laboratory of Mechanics in Energy Engineering, Shanghai University, Shanghai 200072, China}

\author{Lei Kang}
\affiliation{Shanghai Institute of Applied Mathematics and Mechanics, School of Mechanics and Engineering Science, Shanghai Frontier Science Center of Mechanoinformatics, Shanghai Key Laboratory of Mechanics in Energy Engineering, Shanghai University, Shanghai 200072, China}

\author{Yancong Zhang}
\affiliation{Shanghai Institute of Applied Mathematics and Mechanics, School of Mechanics and Engineering Science, Shanghai Frontier Science Center of Mechanoinformatics, Shanghai Key Laboratory of Mechanics in Energy Engineering, Shanghai University, Shanghai 200072, China}

\author{Guohui Hu}
\email{ghhu@staff.shu.edu.cn}
\affiliation{Shanghai Institute of Applied Mathematics and Mechanics, School of Mechanics and Engineering Science, Shanghai Frontier Science Center of Mechanoinformatics, Shanghai Key Laboratory of Mechanics in Energy Engineering, Shanghai University, Shanghai 200072, China}

\title[An \textsf{achemso} demo]
{Piezo1 Decodes Mechanical Forces via Allosteric Network Reprogramming}

\abbreviations{IR,NMR,UV}

\begin{document}

\begin{tocentry}
\centering
\includegraphics{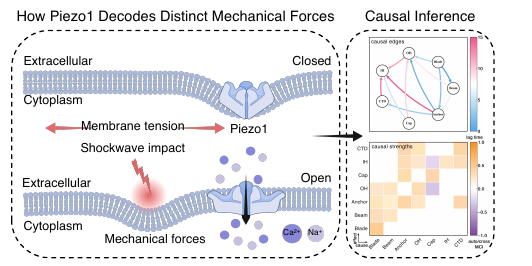}
\end{tocentry}

\begin{abstract}
Understanding how molecular machines transduce mechanical force into chemical signals is a central goal in chemistry. The mechanosensitive ion channel Piezo1 is an archetypal nanoscale mechanotransducer, but the molecular principles by which it decodes distinct mechanical stimuli remain elusive. Here, we combine large-scale molecular dynamics simulations with time-series causal inference to elucidate the dynamic allosteric communication networks within Piezo1 under both quasi-static membrane tension and shockwave-induced cavitation. Under tangential tension, Piezo1 employs the lever-like pathway, a linear, feed-forward pathway propagating the signal from peripheral mechanophores to the central pore. In contrast, a shockwave impulse in the normal direction triggers a two-stage gating mechanism based on the dynamic reprogramming of the allosteric network. An initial compression phase activates an apical shortcut pathway originating from the cap domain. A subsequent tension phase utilizes a rewired network with complex feedback loops to drive the channel to a fully open state. These findings reveal that the allosteric wiring of a molecular machine is not static but can be dynamically reconfigured by the nature of the physical input. This principle of force-dependent pathway selection offers a new framework for understanding mechanochemistry and for designing programmable, stimuli-responsive molecular systems.
\end{abstract}

\section{Introduction}
As a fundamental concept in chemistry and biology, allosteric regulation allows molecular machines like proteins to delicately convert a local input signal into a remote functional output   \cite{BOWERMAN2016429, WOS:000555843700001, Ribeiro2016}.While allosteric events triggered by chemical ligands have been widely documented  \cite{NUSSINOV2013293, Goodey2008, WODAK2019566}, how molecular machines respond to and decode physical inputs of different natures -- for instance, how they distinguish between a sustained quasi-static force and a transient high-energy impulse -- remains largely a challenging problem  \cite{GUO2024102327, Bustamante2004, WOS:001423848900001}. Answering this question is profoundly important to understand the fundamental principles of mechanochemistry \cite{https://doi.org/10.1002/advs.202403949, https://doi.org/10.1002/chin.200542291} and for the optimized design of programmable, intelligent molecular systems  \cite{Li2015, Erbas-Cakmak2015}. 

The allosteric phenomenon has been extensively studied since 1960's. Early landmark models, such as the Monod-Wyman-Changeux (MWC) concerted model  \cite{MONOD196588, annurev:/content/journals/10.1146/annurev-biophys-050511-102222, doi:10.1126/science.1108595} and Koshland-Némethy-Filmer (KNF) sequential,
induced-fit mechanism  \cite{Koshland1966}, described allosterism as the switching of proteins between a few discrete conformational states (e.g., ``relaxed'' and ``constrained'' states). Presently the theory of allosterism has been developed toward a dynamic perspective based on conformational ensembles and energy landscapes \cite{Motlagh2014, https://doi.org/10.1110/ps.03259908, WODAK2019566, 10.1371/journal.pcbi.1003394}. Within this framework, allosterism is no longer regarded as a simple ``switching mechanism," but rather as a ``population shift" process where external perturbations (such as ligands or forces) reshape the energy landscape, leading to the migration of  the population of protein conformations to energetically favored energy well  \cite{doi:10.1073/pnas.041614298, doi:10.1073/pnas.0700329104, KERN2003748}. The kinetic pathway of this process can be described by either ``conformational selection," where pre-existing active conformations are selectively bound, or ``induced fit," where ligand binding induces conformational changes \cite{Vogt2012, doi:10.1073/pnas.0907195106}. This modern theory of allosterism provides a more flexible and profound physicochemical foundation for investigating how a single protein responds to diverse input signals  \cite{NUSSINOV2013293, Changeux2013}.

The mechanosensitive ion channel Piezo1 provides an well-suited model system to address this fundamental question. As a massive, trimeric nanomachine, its function is mediated by mechanical forces exerted by its surrounding lipid membrane environment \cite{doi:10.1126/science.1193270, Saotome2018}. As a molecular mechanotransducer,  Piezo1 is capable of converting a wide spectrum of physical signals into electrochemical signals across diverse cellular contexts \cite{Yang2022, guo2017structure}. For instance, it modulates the responds of the  endothelial cells sense fluid shear stress in the vascular system \cite{ranade2014piezo1}, senses skeletal loads applying on bone cells to help remodeling \cite{li2019stimulation}, participates in the perception of touch and pain in the nervous system \cite{zeng2018piezos}, and regulates the mechanotransduction of soft matrix viscoelasticity \cite{A.G.Oliva2025}. Moreover, its dysfunction, usually related to aberrant activation, might involved in the processes like cancer cell migration \cite{yang2014piezo1}. Its remarkable functional diversity lies in, moderate activation is beneficial while excessive activation could be detrimental \cite{10.3389/fmolb.2025.1693456}. This raises a intriguing question: how does the same molecular entity process different types or magnitudes of mechanical force to produce such distinct, context-dependent outputs? A thorough investigation of its internal allosteric network is crucial to understand this complex phenomenon.

To reveal the allosteric mechanism of Piezo1 in response to quasi-static membrane tension, the mechanochemical coupling has been extensively studied \cite{guo2017structure, young2022physics, yang2023membrane}. In its resting, low-energy state, the lipid membrane of Piezo1 forms a dome-like curvature, maintaining a force equilibrium between the protein and the bilayer \cite{guo2017structure}. Under quasi-static tension, this equilibrium is broken with the membrane flattening and its thickness decreasing. This triggers the conformational cascade starting from the peripheral blade domains. The deformation of blades pulls the beam domain which acts as a molecular lever, propagates the mechanical signal to the inner and outer helix \cite{zhao2018structure}. This allosteric transduction leads to the opening of lateral ion gates on the intracellular side through a ``plug-and-latch" mechanism. Simultaneously, the rotation of the cap domain opens an upper hydrophobic gate, allowing the ion permeation  \cite{WOS:001430059400001}. This theoretical model is widely accepted for the description of a quasi-equilibrium process. It remains an fundamental question in mechanochemistry how the allosteric network of a molecular machine behaves under a high-energy, non-equilibrium impulse (such as a shockwave impact), or whether such extreme perturbation activates novel communication pathways.

The traditional phenomenological analysis is difficult to visualize and analyze these stimulus-dependent dynamic allosteric networks directly. Although conventional molecular dynamics simulations can capture conformational changes, they are usually insufficient to resolve the complex, spatio-temporal causal chains of intramolecular communication \cite{HOLLINGSWORTH20181129, 10.1371/journal.pcbi.1004580}. Causal inference methods, which was developed for complex systems like climate science and neuroscience, offer a powerful framework for this challenge \cite{runge2019inferring, silfwerbrand2024directed}. Specifically, the PCMCI (Peter and Clark Momentary Conditional Independence) algorithm has been successfully identified time-lagged, directional relationships in multivariate, nonlinear systems \cite{runge2019detecting}. By introducing this approach to the study of protein dynamics, the flow of information can be described through a molecular machine's allosteric network. In this study, we first validate our integrated simulation-causal inference methodology by considering the widely-accepted, near-equilibrium allosteric network of Piezo1 under quasi-static tension. Then we apply this approach to investigate the high-energy, non-equilibrium allosteric network driven by shockwave-induced cavitation. By analyzing how Piezo1 decodes the different mechanical signals, we aim to elucidate the fundamental principles of the allosteric network reprogramming in the  molecular machine. 

\section{Materials and Methods}
\subsection{Piezo1 Model Construction and System Setup}
The initial coordinates for the trimeric mouse Piezo1 channel are derived from the cryo-EM structure (PDB: 6B3R) \cite{guo2017structure}. To generate a complete model for simulation, missing residues are added via homology modeling using Modeller (v10.4) \cite{vsali1993comparative}. A topological knot present in the cryo-EM template between residues 2066-2074 is computationally resolved using loop optimization tools, and the optimal loop conformation is selected from 20 candidates based on the DOPE (Discrete Optimized Protein Energy) score \cite{fiser2000modeling, shen2006statistical}. The final construct used for simulations excludes the N-terminal 576 residues and several large, flexible cytoplasmic loops (residues 718-781, 1366-1492, 1579-1654, and 1808-1951). Consequently, each protomer in our model consistes of five non-overlapping polypeptide fragments: 577-717, 782-1365, 1493-1578, 1655-1807, and 1952-2547 \cite{de2021molecular}.

The refined Piezo1 model is then embedded in a 1-palmitoyl-2-oleoyl-sn-glycero-3-phosphocholine (POPC) lipid bilayer using the INSANE script \cite{wassenaar2015computational}. Each system is subsequently solvated and neutralized with 0.15 M NaCl. Two distinct systems are prepared for the different mechanical perturbations: (1) a cubic system for membrane tension simulations, containing 2452 POPC lipids and approximately 179,756 water molecules within an initial 30 × 30 × 30 nm³ simulation box, and (2) an elongated system for shockwave simulations, containing 2452 POPC lipids and approximately 981,078 water molecules in a $30 \times 30 \times 140 \text{nm}^3$ box.

\subsection{Molecular Dynamics Simulation Protocol}
All molecular dynamics (MD) simulations are conducted using the GROMACS 2020.7 software package \cite{abraham2015gromacs}. A hybrid force field scheme is utilized to achieve a balance between computational efficiency and molecular detail. The Piezo1 protein is modeled by the PACE united-atom (UA) force field \cite{de2013improved}, while the POPC lipids, water, and NaCl ions are described with the Martini 2.2 coarse-grained (CG) force field \cite{han2010pace}.

Each assembled system is first subjected to energy minimization by the steepest descent algorithm to eliminate any steric clashes. Following a 500 ps pre-equilibration, a 200 ns equilibration simulation is performed for each system in the isothermal-isobaric (NPT) ensemble with a 2 fs integration time step. During this phase, the temperature is maintained at 310 K using the V-rescale thermostat ($\tau_T = 1.0$ ps) , and the pressure is isotropically controlled at 1 bar using the Parrinello-Rahman barostat ($\tau_P = 1.0$ ps, compressibility = $4.5 \times 10^{-5} \text{bar}^{-1}$) \cite{bussi2007canonical, parrinello1981polymorphic}. No positional restraints are applied during the equilibration run. A cutoff distance of 1.2 nm is used for non-bonded interactions, and periodic boundary conditions are applied in all three dimensions. Convergence to equilibrium is confirmed by monitoring the time evolution of the protein's root-mean-square deviation (RMSD) and radius of gyration (Rg). The final equilibrated box dimensions are 29.4 × 29.4 × 29.6 nm³ for the stretch system and 29.6 × 29.6 × 138.5 nm³ for the shockwave system.

\subsection{Membrane Stretch and Shockwave}
After equilibration, two different non-equilibrium simulation protocols are considered to study the mechanochemical response of Piezo1.

{\bf Membrane Stretch Simulations.} To model the channel's response to quasi-static tension, a constant lateral pressure of -40~bar is applied to the $xy$-plane of the system. These production simulations are performed in the NPT ensemble. The integration time step is set to 2 fs.

{\bf Shockwave Simulations.} The channel's response to a high-energy impulse is investigated by generating a planar shockwave using the momentum mirror method \cite{choubey2011poration}. A rigid wall composed of carbon atoms (25 atoms/nm³) is placed at one end of the elongated simulation box, separated from the system by a 2 nm vacuum layer. To model cavitation-induced shock, a spherical void (nanobubble) with a 10 nm radius is created by removing water molecules 5 nm away from the membrane surface. All particles in the system are initialized with a constant velocity ($u_p$) directed towards the mirror wall. Upon collision, the wall reverses the $z$-component of a particle's velocity, launching a shockwave that propagated back through the system towards the protein-membrane assembly (Figure 1a). These shockwave simulations are conducted in the microcanonical (NVE) ensemble for 60 ps with a 1 fs time step. Periodic boundary conditions along the $z$-axis (the direction of shock propagation) are disabled.

\subsection{Statistics and Causal Inference}
\begin{figure}
	\centering
	\includegraphics[scale=0.6]{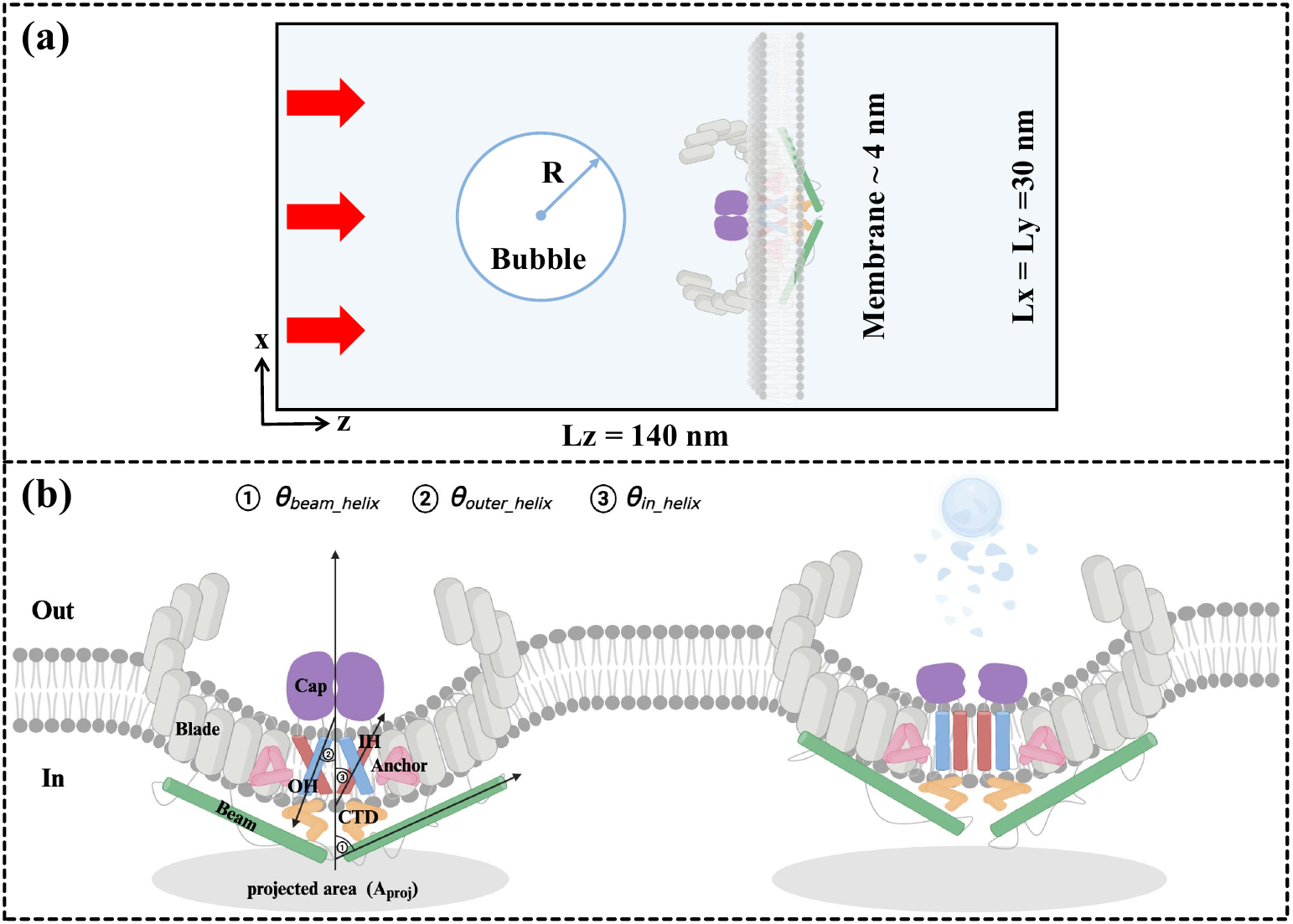}
	\caption{Schematic of shockwave impulse. (a)A planar shockwave generated by the particle propagates to the right, causing bubble collapse and generating a jet that impinges on the membrane protein. (b) Resting structure of Piezo1 and its activation driven by shockwave impulse. Shown are the projected area of the protein and the tilt angles of key helices (Beam, OH, IH) relative to the membrane normal.}
	\label{fg:1}
\end{figure}
{\bf Input Data Preprocessing.} To map the allosteric communication pathways, causal inference is performed on the time series of the root-mean-square deviation (RMSD) for each protein domain, which serves as a proxy for their conformational dynamics. To isolate the high-frequency fluctuations relevant for signal propagation from the overall conformational drift, the raw RMSD time series are decomposed using the Seasonal-Trend-Loess (STL) method \cite{rb1990stl}. By standardizing, the residual components with zero mean and unit variance are obtained. These time series (see Supplementary Figures S1-S2) are used as the input for the causal network analysis.

{\bf Causal Network Inference.} Causal networks are constructed using the PCMCI (Peter and Clark Momentary Conditional Independence) algorithm within the Tigramite Python toolbox \cite{runge2019inferring}. The PCMCI method is particularly well-suited for identifying directional, time-lagged relationships in complex, nonlinear systems, making it ideal for tracking intramolecular signal propagation. The Robust Partial Correlation (RobustParCorr) test is employed as the conditional independence criterion. The maximum time lag $\tau_{max}$, which defines the temporal window for detecting causal links, is determined for each simulation by analyzing the decay of the time series autocorrelation functions (see Supplementary Figures S3, S8, S10). The robustness of the resulting network topology to the choice of $\tau_{max}$ has been verified through sensitivity analyses.

{\bf Statistical Validation and Network Construction.} A statistical significance level of $\alpha = 0.05$ is used for all conditional independence tests. To control the false discovery rate (FDR), the p-values of all potential links are adjusted using the Benjamini-Hochberg procedure \cite{runge2018causal}. Only the causal links that remains significant after this correction are kept to construct the final directed graphs, which represent the statistically validated allosteric communication networks within the Piezo1 channel.

\section{Results}
\subsection{Quasi-Static Membrane Tension: A Benchmark}
\subsubsection{Conformational Cascade}
\begin{figure}
	\centering
	\includegraphics[scale=0.35]{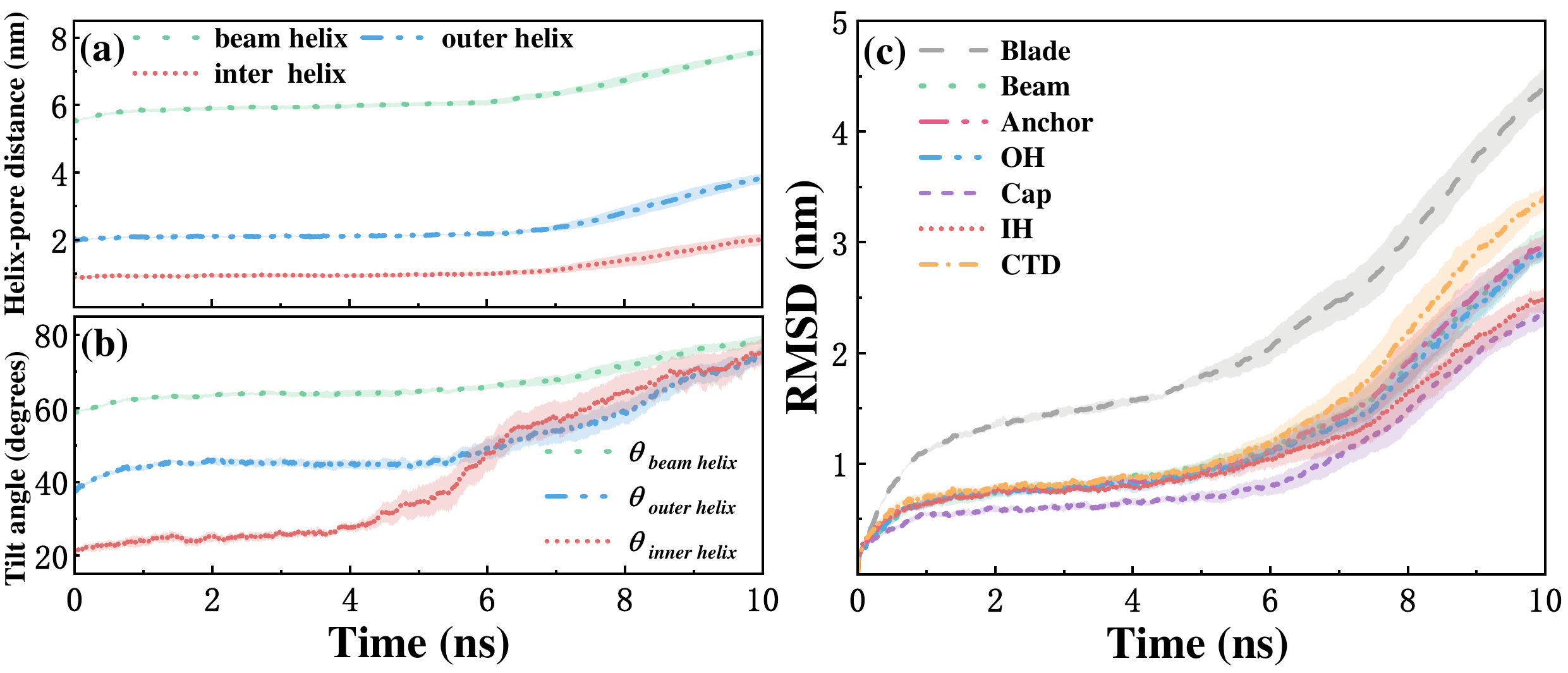}
	\caption{Conformation variation in Piezo1 under membrane tension. (a) Temporal evolution of the distance between the key helix and the gating center of mass. (b) Temporal evolution of the angle between the key helix and the membrane normal. (c) RMSD of the structural domains as a function of time.}
	\label{fg:2}
\end{figure}
To validate our integrated simulation-inference methodology, we try to obtain the allosteric communication network of Piezo1 under near-equilibrium condition -- quasi-static membrane tension. A constant lateral pressure of -40~bar is applied to the lipid bilayer to induce gating-related conformational changes on an affordable simulation timescale in previous study \cite{de2021molecular}. It is worthy to note that this supra-physiological tension serves as a computational tool to accelerate the conformational transition, rather than representing a physiological condition in reality.

Our simulations reveals the mechanochemical coupling between the membrane and the ionic channel. Under the applied tension, the lipid bilayer continuously thins, and the projected area of the Piezo1 expands in the meantime. This geometric rearrangement initiates a coordinated conformational cascade in the force-transducing helices of the channel. A continuous and significant radial displacement away from the central pore's center of mass is observed for the beam helix, outer helix (OH), and inner helix (IH) (Figure 2a). This outward movement is simultaneously coupled with an increase in their tilt angles relative to the membrane normal direction (as defined in Figure 1b; see Figure 2b), suggesting its reorientation toward the membrane plane. This concerted motion -- a combination of radial expansion and angular tilting --  results in the dilation of the central pore diameter. Within the scope of this study, this structural variation is used as the  indicator of the channel transitioning to a gate-open conformation.

To elucidate the molecular mechanism underlying the observed pore dilation, our goal is to move beyond describing conformational changes and to find the directional flow of information within the Piezo1 molecular machine. The conformational response of each domain is first quantified by analyzing the time evolution of their root-mean-square deviations (RMSD) depicted in Figure 2c. The results show that the peripheral Blade domain undergoes the earliest and most substantial conformational rearrangement, due to the largest RMSD increase during the simulation. This observation provides the dynamic evidence that the Blades function as the primary mechanophore, directly converting membrane strain into a large-scale structural change. The other domains, including the Beam, Anchor, and pore-lining helices, also displayed significant RMSD increases but with a temporal delay relative to the Blades, implying a sequential propagation of the conformational change from the periphery toward the central pore.

\subsubsection{Allosteric Communication Network under membrane tension}
\begin{figure}
	\centering
	\includegraphics[scale=0.35]{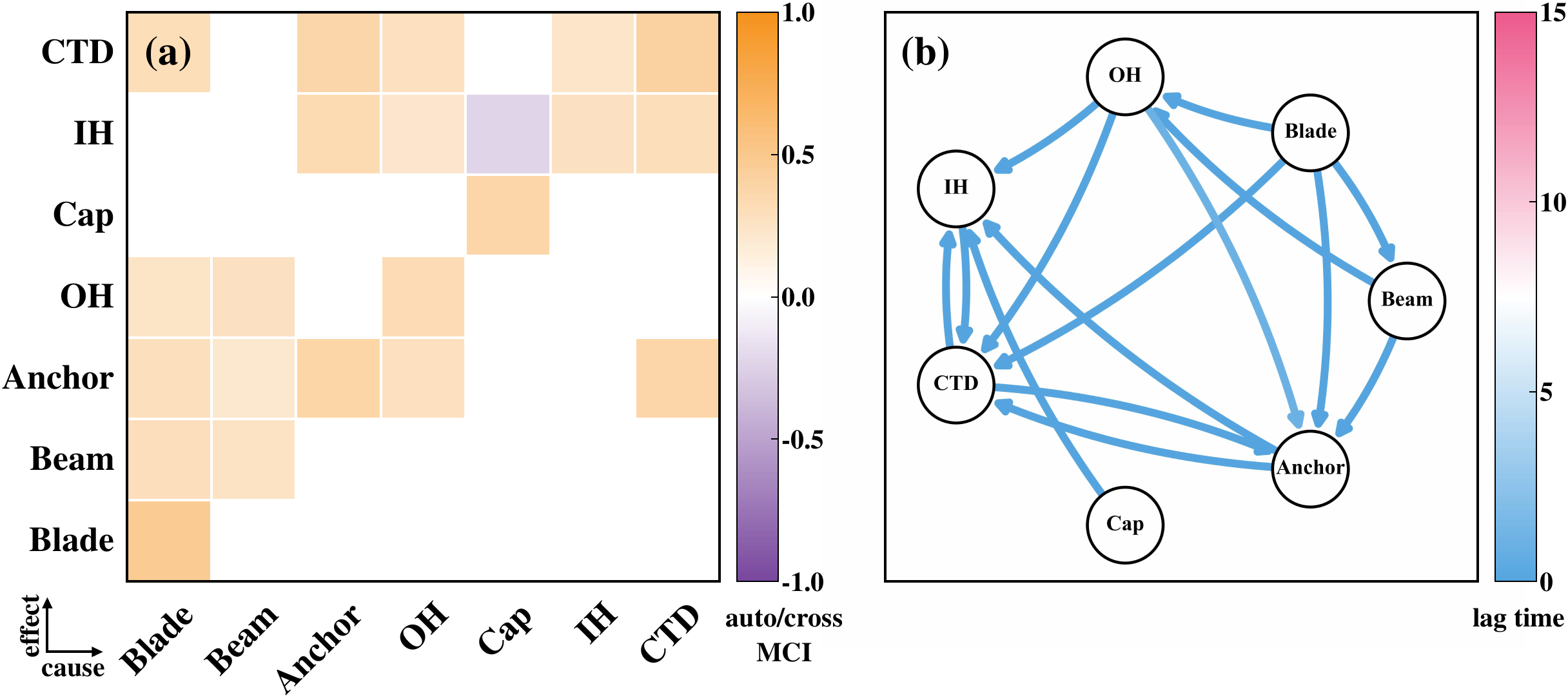}
	\caption{Allosteric communication network of Piezo1 under membrane tension. (a) Autocorrelation (auto-MCI) and cross-domain causal interaction (cross-MCI) strengths among Piezo1 structural domains. Orange indicates positive correlation, purple indicates negative correlation, and the color intensity reflects the interaction strength. Auto-MCI quantifies the influence of a domain’s own history on its current state, whereas cross-MCI captures the influence of other domains’ histories on the current state. (b) Causal network of Piezo1 structural domains. Directed edges point from cause to effect, and the color of each arrow denotes the corresponding lag time.}
	\label{fg:3}
\end{figure}
To identify the origin of this signaling cascade, a causal network is constructed via causal inference analyses. The determination of the maximum lag time $\tau_{\max}$ and the robustness checks of the inferred causal relationships are provided in Supplementary Information, Figures S3–S4. It confirms that the Blade domain is indeed the principal initiator, with the highest auto-causality (Figure 3a, auto-MCI = 0.467, Table S1 in Supplemental Information), which indicates that the Blades possess the strongest autonomous conformational response to the applied membrane strain, as the primary mechano-sensors of the system. This finding is consistent with previous studies  \cite{zhao2018structure} identifying the Blade domain as the initial sensor of membrane tension.

The causal network illuminates that the Blade domain acts as the origin of the network, exerting nearly simultaneous causal influences on the Beam, Anchor, CTD, OH, and IH, as shown in Figure 3b. This indicates that the initial perception of membrane tension is primarily undertaken by the Blade, then transmits in parallel along multiple pathways to the core domain. Among all candidate paths, the Blade → Beam → Anchor → CTD → IH sequence represents a dominant and complete conduction chain. The Blade drives the rotation of the Beam, which acts as a molecular lever in force transmission, amplifying and relaying the external tension to the Anchor. The Anchor subsequently serves as an interface, acting upon the CTD to ultimately regulate the IH. This result provides quantitative evidence to confirm the phenomenological description proposed by Wang et al.  \cite{wang2018lever}.

In addition to this main pathway, multiple auxiliary pathways are also found in the network of information flows, such as Blade → Beam → Anchor → IH and Blade → Beam → OH → IH. These shortcuts or parallel paths suggest that the gating of Piezo1 does not rely on a single chain but rather on the concerted action of multiple pathways to enhance the system's robustness. Meanwhile, the Blade → OH → IH side path implies that the OH domain may be directly involved in gating regulation in certain activation states, adding diversity to the response patterns.

Furthermore, the network is capable of identifying several crucial nexus and feedback relationships that are difficult to describe using purely phenomenological methods. As an information nexus within the network, the CTD connects with the Anchor by a strong bidirectional interaction, although they are not structurally adjacent. These dynamic coupling suggests they may form a platform for signal integration and redistribution in the near-pore region, potentially mediated by the membrane environment or the concerted motion of neighboring helices. 
The OH not only receives inputs from the Blade and Beam, but also acts upon the Anchor, CTD, and IH, playing a role in distributing and coordinating signals during lateral conduction. As a core effector domain for signal integration, the IH is regulated by multiple pathways and can, in turn, influence the CTD, suggesting the existence of a feedback loop to maintain stability or to fine-tune the gating process. 

A particularly intriguing finding is the negative causal link between the Cap and IH domains. This link should be interpreted not as a simple suppressive signal, but rather as the signature of an inverse coupling relationship inherent to the channel's mechanical design. The initial outward displacement of the Blades disrupts a pre-existing mechanical equilibrium between the peripheral and apical structures. This disruption necessitates a compensatory motion in the Cap domain, which is physically coupled to the outward movement of the IH required for pore dilation. This finding suggests that the top structures participate in gating through a sophisticated mechanical mechanism governed by the principles of coupled motion, rather than a simple, linear signaling cascade.

In summary, the present causal inference analysis has successfully uncovered the near-equilibrium allosteric network that governs Piezo1's response to quasi-static tension. The network topology is dominated by a primary, feed-forward cascade from the peripheral Blade mechanophores to the central pore modulus, which is consistent with the classic lever-like mechanism. Our approach, however, also resolves a richer landscape of parallel pathways, feedback loops, and signaling nexus points, revealing a more complete picture of the intramolecular communication. The successful elucidation of this complex network provides a robust validation of our integrated methodology. We now apply this validated approach to investigate a fundamentally different regime -- the non-equilibrium allosteric response of Piezo1 to a high-energy shockwave impulse.

\subsection{Non-Equilibrium Gating Mechanism under Shockwave Impacting}
\subsubsection{Two-Stage Gating Pathway: From Compression to Tension}
\begin{figure}
	\centering
	\includegraphics[scale=0.35]{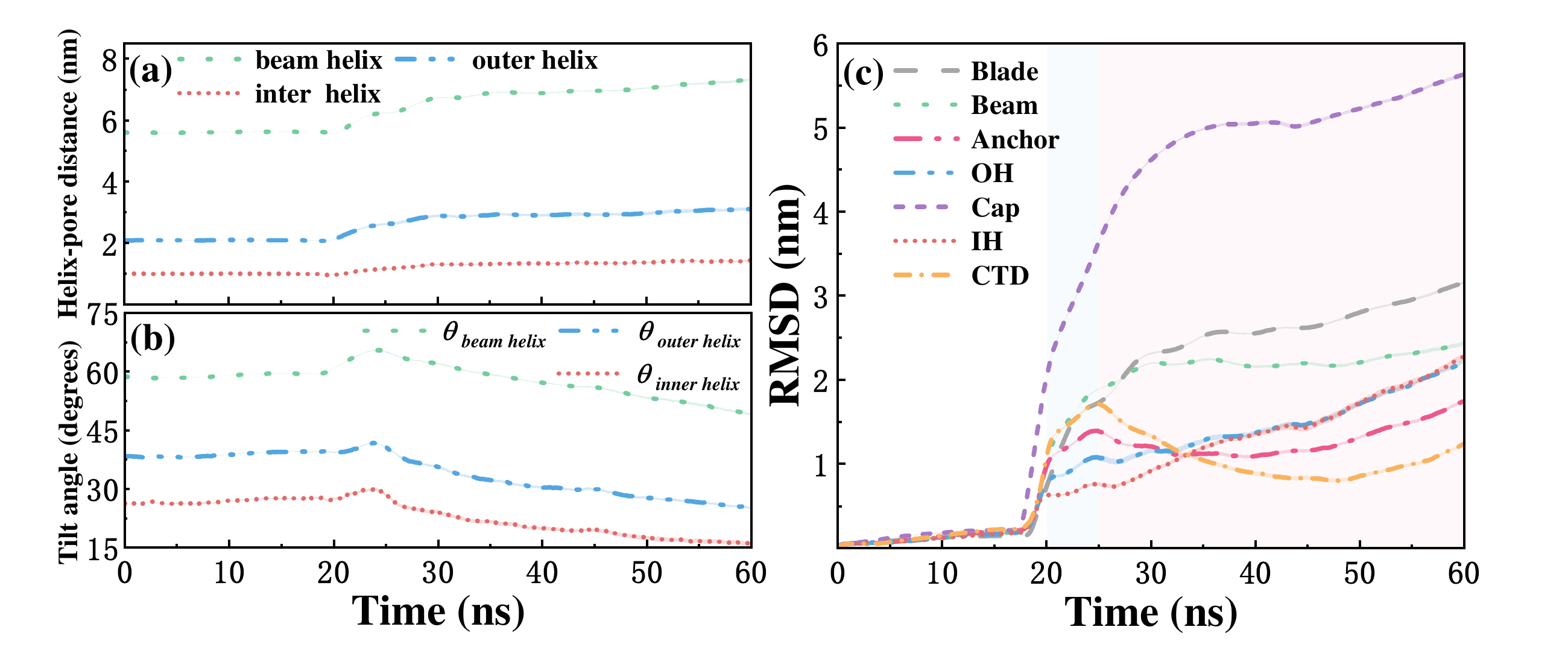}
	\caption{Conformation variation in Piezo1 under shockwave impulse. (a) Temporal evolution of the distance between the key helix and the gating center of mass. (b) Temporal evolution of the angle between the key helix and the membrane normal. (c) RMSD of the structural domains as a function of time.}
	\label{fg:4}
\end{figure}
Before analyzing the detailed gating pathway,  the non-equilibrium simulation protocol is performed to reproduce the well-established requirement of cavitation for shockwave-induced channel activation. Consistent with previous studies \cite{Nan2018nanoscale}, it is found that a pure shockwave (in the absence of a nanobubble) is insufficient to cause significant conformational rearrangement or gating across the tested range of impact velocities. In contrast, the presence of a cavitating nanobubble is essential to induce the large-scale protein dynamics and radial expansion in the pore, leading to a conductive open state (Figure S5 in Supplementary Information). This result confirms the critical role of cavitation as the mechanochemical trigger and validates the physical fidelity of our simulation setup. Therefore, all subsequent analyses focus  on exploring the mechanism of cavitation-induced gating.

We selected a particle velocity of $\mathrm{u_{p}= 1.0 km/s}$ for a detailed analysis of the dynamic pathway, as this condition is known to generate shockwaves consistent with experimental results (see Figure S6)\cite{choubey2011poration, adhikari2016nanobubbles} and induced the most significant conformational changes in our initial tests. 

After an initial ~20 ps lag for the shockwave to reach the protein, the key force-transducing helices behaviors a unique, biphasic motion. While their radial distance from the pore center continuously increased after 20 ps (Figure 4a), their tilt angles showed a non-monotonic evolution, i.e., a rapid increase between 20-25 ps, followed by a sustained decrease from 25-60 ps (Figure 4b). This non-monotonic angular reorientation is the solid evidence of a ``two-stage synergistic'' gating mechanism driven by distinct mechanical modes.

The first stage (20-25 ps) corresponds to a compression-driven pre-activation. The primary force during this phase is the axial compressive impulse from the jet produced by the nanobubble collapse. This impulse forces, acting upon the protein, lead to a radial expansion of the entire propeller-like assembly. This mechanical signal propagates down the Blade-Beam allosteric cascade, driving the key helices to tilt further away from the membrane normal (angle increase in Figure 4b). This conformational change, while resulting in preliminary pore expansion, more importantly drives the protein into a ``pre-activated" intermediate state, likely enhancing its sensitivity to the subsequent mechanical perturbation.

The second stage (25-60 ps) witnesses a tension-driven full opening. The dominant force transitions to a more sustained axial tensile stress, created by the fluid dynamics following the jet impact. This axial tension alters the global mechanical equilibrium of the channel, initiating a conformational relaxation where the key helices realign toward the membrane normal (angle decrease). This realignment is simultaneously coupled with continued radial expansion (distance increase). It is this concerted motion -- the combination of axial realignment and radial expansion -- that ultimately leads to the dilation of the pore to its fully open and conductive state. Overall, the gating of Piezo1 under a shockwave impulse is not a single-step event, but a sequential, coupled process, revealing a unique mechanochemical pathway different from its near-equilibrium counterpart. 

\subsubsection{Reprogramming of the Allosteric Network under a Shockwave Impulse}
\begin{figure}
	\centering
	\includegraphics[scale=0.35]{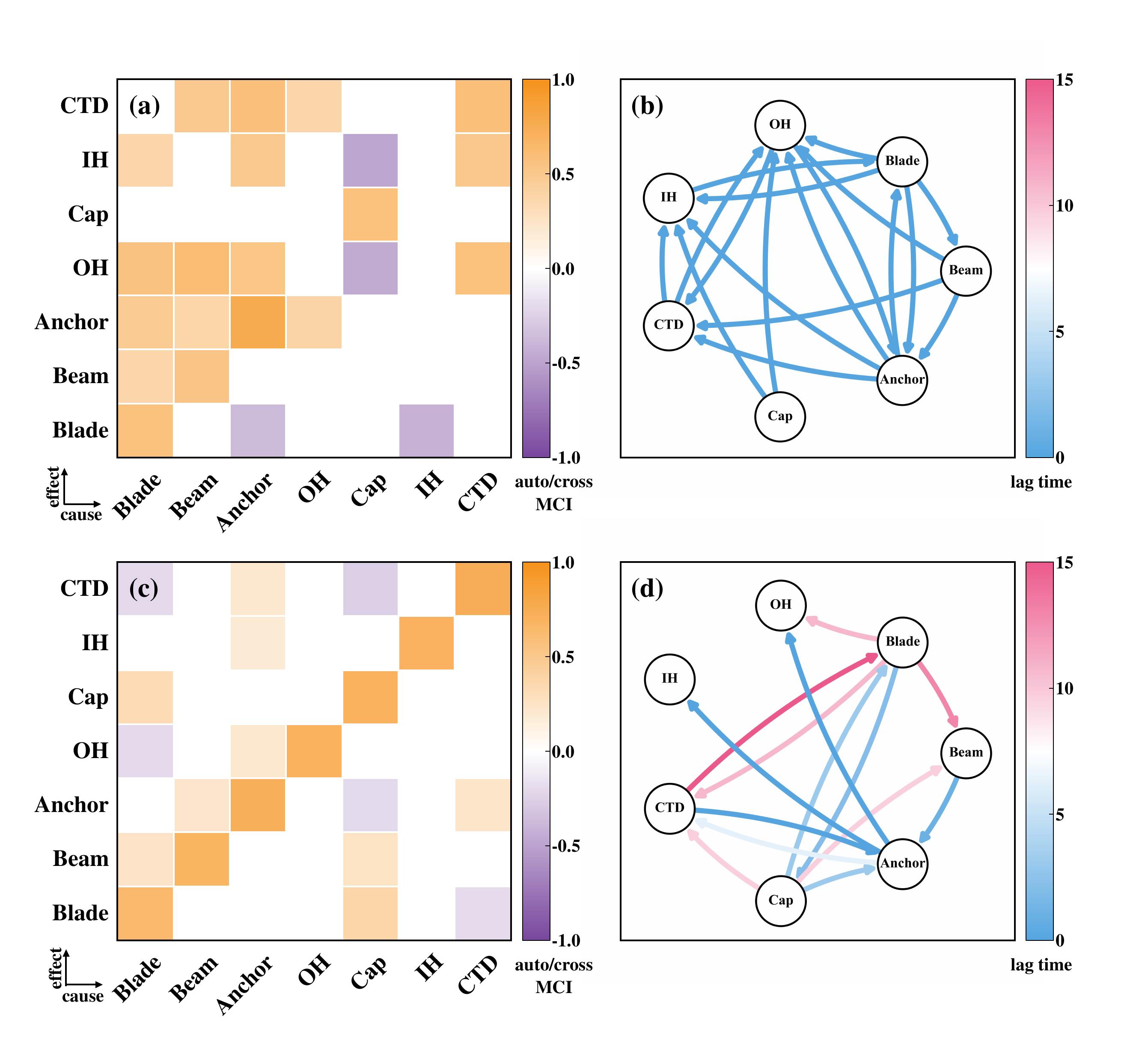}
	\caption{Allosteric communication network of Piezo1 under shockwave impulse. (a) and (b) correspond to the first phase of the shockwave, whereas (c) and (d) correspond to the second phase. (a) and (c) Autocorrelation (auto-MCI) and cross-domain causal interaction (cross-MCI) strengths among Piezo1 structural domains. Orange indicates positive correlation, purple indicates negative correlation, and the color intensity reflects the interaction strength. (b)and (d) Causal network of Piezo1 structural domains. Directed edges point from cause to effect, and the color of each arrow denotes the corresponding lag time.}
	\label{fg:5}
\end{figure}
To quantitatively characterize the conformational dynamics of each domain under the shockwave impulse, we further analyze the time evolution of their respective RMSD. By examining the first derivative of the STL-decomposed trend of each RMSD time series, a significant extremum in the derivatives can be found at approximately 25 ps, corresponding to a critical transition in the conformational change (see Figure S7 in Supplementary Information). This transition is consistent with the moment when the system's response shifts from being compression-dominated to tension-dominated as descibed in Figure 4b. Therefore, it is necessary to perform causal inference analysis on two stages separately to construct the allosteric communication networks under these distinct mechanical regimes.

As illustrated by the RMSD evolution profiles in Figure 4c, apparently the domains under the shockwave impulse behaviors differently from their response under membrane tension. During the initial 0–20 ps period, before the shockwave front reaches the protein, all domains remain at low RMSD values, indicating conformational stability. The compression phase (20–25 ps) is initiated by the jet from the collapsing nanobubble, triggering a near-synchronous, profound increase in the RMSD of multiple domains, albeit with distinct magnitudes. The Cap domain undergoes a violent conformational perturbation, and its RMSD raises far beyond that of any other domain. Since the Cap is the most outwardly exposed structure encountering the shockwave front firstly, this dramatic perturbation suggests that it functions as a primary sensing module in this transient mechanical regime. The RMSD variations of the Blade, Beam, and CTD follow immediately, implying the mechanical signal propagates from the periphery domain. Subsequently the Anchor is disturbed, acting as a signal relay and amplification hub before the signal propagates to the OH and IH domains.

Upon entering the tension-dominated phase (25–60 ps), the system transitions into a process of conformational relaxation and reorganization that unfolds over a significantly longer timescale. The Cap's RMSD remains the highest and continues to increase, indicating that it maintains a state of high conformational mobility. The Blade sustains a strong response, continuously delivering mechanical information inward. In contrast, the Beam's RMSD gradually reaches a plateau, because it has completed its major reorientation and settled into a stable force transmission state. The CTD and Anchor display a characteristic fall-then-rise pattern, which can be interpreted as an phase of stress release or relaxation, followed by their re-engagement into the primary force-transduction chain. Concurrently, the OH and IH domains show a monotonic increase in RMSD and their values progressively converge. This demonstrates that the OH and IH are not responding in isolation; instead, they experience a synergistic motion involving synchronous reorientation and radial expansion to achieve the fully open state.

To intensively explore the mechanisms of structural coupling under the shockwave, we partition the causal inference analysis into the two dynamic phases identified previously, i.e., the compression phase (20–25 ps) and the tension phase (25–60 ps). Corresponding causal networks are then constructed for each phase to examine the dynamic evolution of the allosteric communication pathways. Importantly, a different maximum lag time $\tau_{\max}$ is determined for each phase, according to their fundamentally different timescales and correlation decay properties (see Supplementary Information, Figures S8-S11).

{\bf Phase I: Allosteric Pathway Activated by Compression (20-25 ps)}. In the high-energy compression phase, the causal network (Figure 5b) reveals a dramatic departure from the near-equilibrium state. The autocorrelation analysis (Table S2) shows that the Anchor domain possesses the highest path dependency (Figure 5a, auto-MCI = 0.766), acting as a stabilizing fulcrum. Conversely, the level of autocorrelation for the pore-lining OH and IH domains is negligible, identifying them as passive recipients of upstream signals. Based on these observations, the network's active pathways are then investigated. The most prominent trunk pathway remains the canonical Blade → Beam → Anchor → CTD → OH/IH cascade, demonstrating the preservation of the outside-in transduction logic. In the meantime, a rapid allosteric route emerges, originating directly from the Cap domain. This apical shortcut pathway (Cap → OH/IH) effectively bypasses the well-known lever system. This finding, which is consistent with the Cap's dramatic RMSD perturbation, establishes it as the primary mechanophore in the non-equilibrium shock impacting. The importance of the Cap domain in Piezo gating has also been examined in previous studies \cite{JIANG2021472, LEWIS2020870, Wang2019}, in which it is found that deleting the whole Cap domain or the region in contact with the Blade, or restricting the Cap motion by crosslinking will abolish its mechanical activation. Furthermore, the network resolves several auxiliary pathways. The Blade → Beam → CTD → IH pathway suggests that the CTD is not passively delayed during the compression phase but rather engages the IH in the core driving mechanism, evidenced by the significant rise in the CTD's RMSD. Meanwhile, the Blade → Beam → Anchor → OH/IH pathway indicates that the Anchor, in this phase, acts not only as the lever-anchor fulcrum but can also directly drive the neighboring OH and IH, allowing them to participate in the gating process before the CTD fully responds. This correlates with the near-synchronous rise observed in the RMSD of both OH and IH. 

{\bf Phase II: Network Reprogramming under Tension (25-60 ps).} As the system transitions into the slower, tension-driven phase, the allosteric network undergoes a significant topological rewiring (Figure 5d). Dramatic variation in the autocorrelation landscape can be observed. The CTD and Anchor have the highest autocorrelation coefficients (Figure 5c, 0.740 and 0.723, respectively), showing their role as a support framework for the protein opening (Table S3). The level of autocorrelation of the OH and IH domains increases significantly, signifying their transition from passive elements into active effector domains with sustained, stable dynamics. The causal network reflects this functional transition. The primary Blade → Beam → Anchor pathway is preserved, but the signal now diverges at the Anchor nexus into two principal branches: one directly targeting the effector domains (Anchor → IH/OH) and another acting on the support framework (Anchor → CTD). Impressively, numerous bidirectional causal links -- i.e., feedback loops -- emerge, a feature absent in the scenarios we discussed before. For instance, strong coupling is observed between the Anchor and CTD, as well as between the Cap and Blade. These feedback mechanisms indicate a functional transition to a more integrated, cooperative process, where key nodes not only receive inputs but also engage in regulatory feedback to realize the fully open conformation. The Cap domain, while no longer the primary driver, remains a key coordinator through its multiple feedback connections to the Blade, Beam, and CTD.

\section{Conclusions}

By integrating large-scale molecular dynamics with time-series causal inference, the present study has systematically explored the allosteric mechanisms governing the gating of the Piezo1 channel under fundamentally different mechanical regimes. Our central finding is that the intramolecular communication network of this sophisticated molecular machine is not a static, hard-wired circuit. Instead, it can be reprogrammed by the specific nature of the mechanical stimulus, a principle we term as force-dependent selection of allosteric pathways.

Under quasi-static membrane tension, our analysis quantitatively validates the operation of the lever-like transduction pathway \cite{wang2018lever}. The network topology in this near-equilibrium state is characterized by a linear, feed-forward architecture, where the signal propagates sequentially from the peripheral Blade mechanophores, through the Beam-Anchor lever-like system, to the central pore. The successful characterization of this pathway provides a robust benchmark, validating the feasibility of our integrated simulation-inference methodology.

Subsequently, it is found that a high-energy shockwave impulse in the normal direction triggers a non-equilibrium two-stage gating mechanism based on the rewiring of the allosteric network. The initial compression phase (20–25 ps) activates the apical shortcut pathway, where the Cap domain, acting as the primary shock-response mechanophore, creates a direct shortcut to the pore-lining helices. This achieves a rapid ``pre-activation" of the channel. Successively, in the tension-driven phase (25–60 ps), the network is further reconfigured to include numerous feedback loops between the key domains, and these emergent feedback mechanisms are critical for realizing the open conformation through a more integrated and cooperative process.

While our simulations provide a dynamic model with deep physical insights, it must be recognized that these findings currently originate from in silico models; therefore, it is crucial to validate these computational predictions with experimental studies. The mechanistic blueprint we proposed provides helpful guidance for future experiments. For example, the technologies such as time-resolved cryo-electron microscopy (Cryo-EM) or X-ray free-electron lasers (XFEL) are promising to directly capture the pre-activated intermediate state we predict. Additionally, by performing site-directed mutagenesis at the key interfaces of the ``apical shortcut pathway'' (e.g., the Cap-OH/IH interface) and combining this with single-channel electrophysiology recordings, the functional contribution of this pathway to shockwave-induced gating might be directly tested. Moreover, advanced spectroscopic techniques, such as Nuclear Magnetic Resonance (NMR) relaxation dispersion and single-molecule FRET (smFRET), can provide quantitative data on the population shifts of conformational ensembles and the kinetic rates of state transitions under different stimuli, thus dynamically validating the occurrence of pathway selection.

Beyond the specific findings for Piezo1, this work represents a significant step towards a new paradigm in the study of molecular biophysics. Traditional approaches to understanding channel gating have often relied on phenomenological descriptions, where the rules of conformational change are qualitatively inferred. Our research, however, introduces a data-driven framework capable of quantitatively deconstructing the complex web of intramolecular cause-and-effect relationships. This attempt to shift the focus from qualitative models to quantitative, time-resolved network analysis offers a powerful tool to understand the sophisticated dynamics of molecular machines.

Recently, the intensive integration of artificial intelligence (AI) and machine learning (ML) is significantly helpful for the progress in biological molecules  \cite{10.1063/1.4998259, BERNETTI2024102820, WOS:000852292300001}. For a giant protein like Piezo1, the timescale of its allosteric process poses an great challenge for conventional MD simulations. Combining machine learning algorithms with enhanced sampling techniques is expected to explore its complete conformational energy landscape with high efficiency, automatically identifying key allosteric intermediate states and transition pathways. Furthermore, deep learning models can learn directly from a protein's sequence, structure, or dynamic data to predict potential allosteric networks and regulatory ``hotspots'', ultimately achieve the leap from ``analysis'' to ``design''. Our findings demonstrate the programmability of internal protein pathways, providing a theoretical basis and possibility for the future de novo design of synthetic allostery proteins with specific mechanical response logic using AI.

In conclusion, this research not only provides a foundational understanding of Piezo1's gating under shockwave conditions, but also introduces causal inference for understanding mechanochemistry. The discovery that a single molecular entity can dynamically select its internal communication pathways based on the temporal profile of a physical force deepens our fundamental knowledge of cellular mechanotransduction. This principle offers a new theoretical basis for optimizing medical applications such as shockwave therapy, suggesting that the precise tuning of physical parameters could be used to selectively engage desired biological pathways, foreshadowing a new era of programmable, stimuli-responsive molecular interventions.

\begin{acknowledgement}

This research is supported by National Natural Science Foundation of China (Nos. 12332016 and 12172209), the Science and Technology Commission of Shanghai Municipality (Grant No. 24TS1412500). Beijing Beilong Supercloud Computing Co., Ltd. provides support in computational resources.

\end{acknowledgement}

\section*{Data Availability}
The data that support the findings of this study are available from the corresponding author upon reasonable request.

\begin{suppinfo}

Details of molecular dynamics simulations, parameter settings, and additional analyses (Figures S1–S11 and Tables S1–S3) are available in the Supporting Information.

\end{suppinfo}

\bibliography{main}

\end{document}


\section*{Table of Contents}

\subsection*{A. Preprocessing of RMSD time series using STL decomposition}

\subsection*{B. Selection of $\tau_{\max}$ and robustness of PCMCI causal networks under membrane tension}

\subsection*{C. Conformational responses of the membrane protein under shockwave impact with and without cavitation}

\subsection*{D. Validation of shockwave modeling}

\subsection*{E. Identification of the compression–tension transition under shockwave impacting}

\subsection*{F. Selection of $\tau_{\max}$ and robustness of PCMCI causal networks during shockwave compression}

\subsection*{G. Selection of $\tau_{\max}$ and robustness of PCMCI causal networks during shockwave tension}

\newpage
\section*{A. Preprocessing of RMSD signals using STL decomposition}
The RMSD time series of the Piezo1 domains are inherently non-stationary, which is difficult for lagged-correlation analysis and PCMCI causal inference to accurately capture the underlying coupling between domains. Therefore, STL (Seasonal–Trend decomposition using LOESS) is applied to preprocess each RMSD time series \cite{rb1990stl}.

STL decomposes each time series into three components: (1) a trend component, representing the overall structural deformation of the system. Under membrane tension, the trend captures a slow structural drift (Figure S1), whereas under shockwave impact it demonstrates a transient and pronounced deformation response (Figure S2); (2) a seasonal component, capturing the periodic or quasi-periodic oscillations arising from membrane undulations, thermal fluctuations, or related motions; (3) a residual component, consisting of the high-frequency, non-periodic fluctuations that remain after removing the trend and seasonal terms. These residuals primarily reflect the intrinsic fast conformational dynamics of each domain and constitute the most informative signal for subsequent dynamical analyses, including the extraction of true inter-domain coupling.

To obtain stationary and comparable inputs for causal inference, $z$-score normalization is applied to the residuals, mapping the signals of different domains onto a unified scale. This normalization removes amplitude differences in RMSD fluctuations across domains, ensuring the residual series faithfully associated with their true coupling strengths. In addition, by removing the trend and seasonal components, the resulting time series closely approximate to stationarity, thereby satisfying the PCMCI requirement for stable input signals.

\begin{figure}[H]
	\centering
	\includegraphics[scale=0.25]{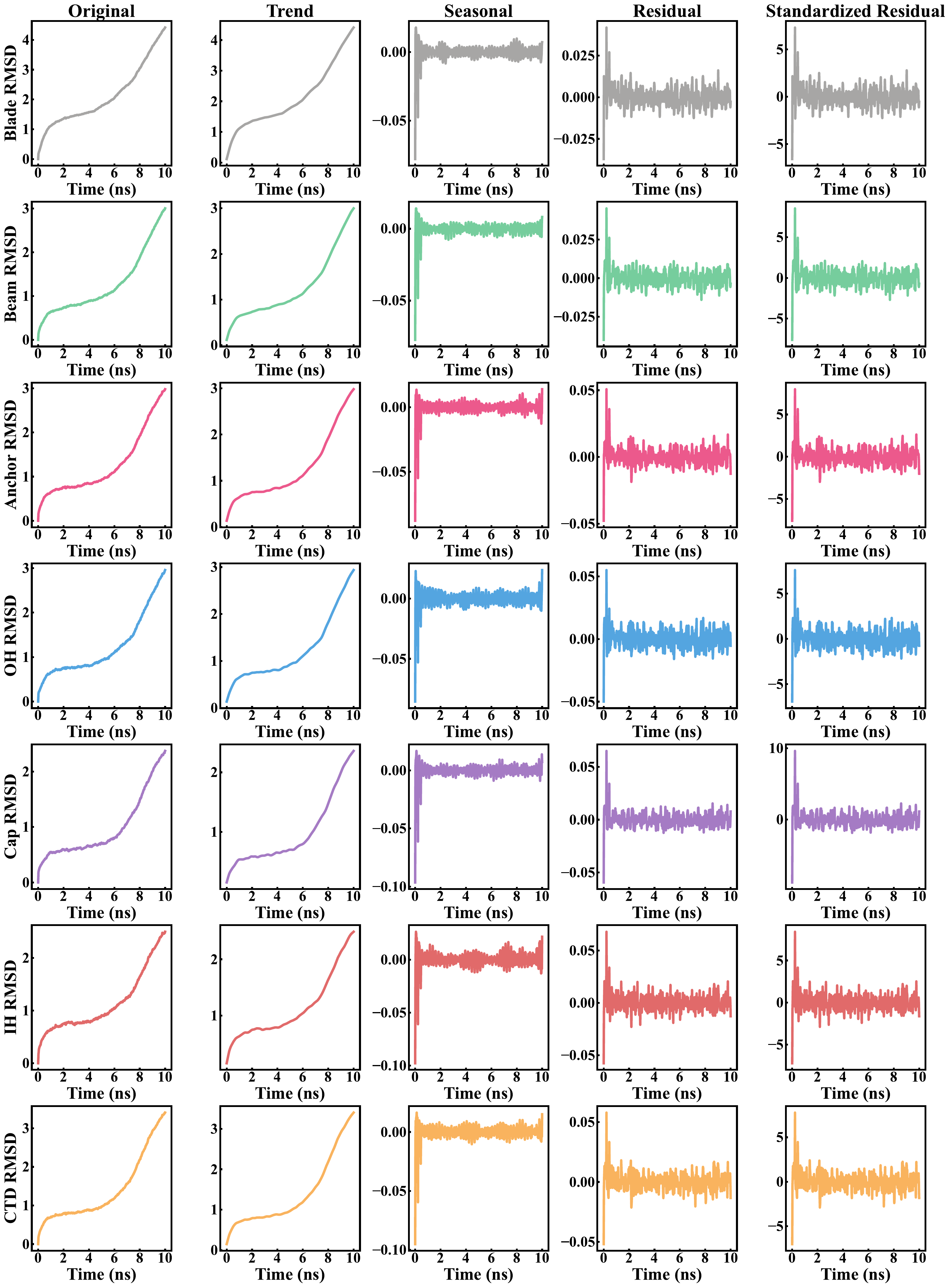}
        \caption*{Figure S1: STL decomposition of RMSD time series for the seven Piezo1 domains under membrane tension. Each panel displays the raw RMSD time series, the trend component, the seasonal component, the residual component, and the standardized residuals used for causal inference.}
	\label{fg:S1}
\end{figure}

\begin{figure}[H]
	\centering
	\includegraphics[scale=0.25]{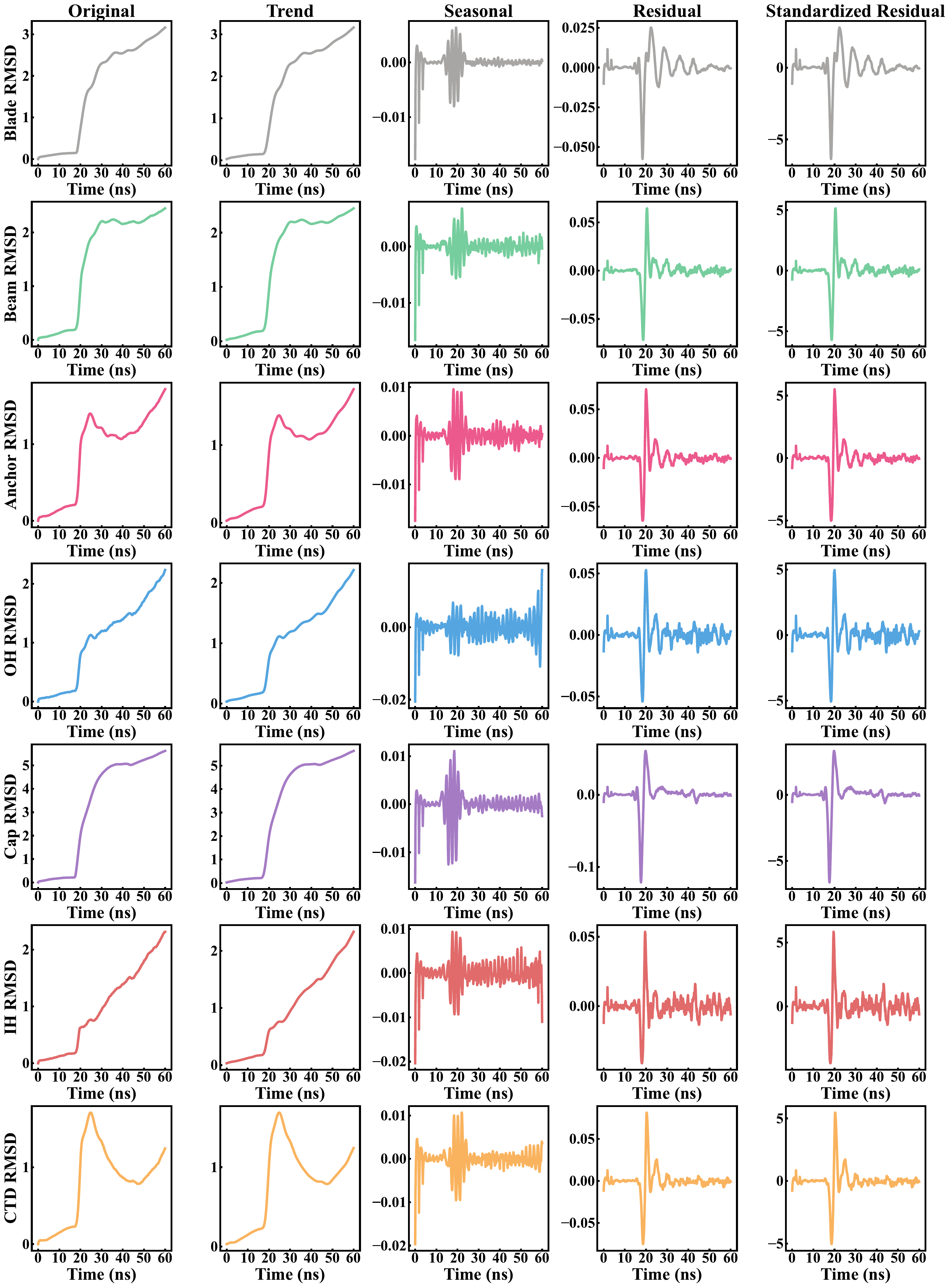}
	\caption*{Figure S2: STL decomposition of RMSD time series for the seven Piezo1 domains under shockwave impact. Each panel displays the raw RMSD time series, the trend component, the seasonal component, the residual component, and the standardized residuals used for causal inference.}
	\label{fg:S2}
\end{figure}

\section*{B. Selection of $\tau_{\max}$ and robustness of PCMCI causal networks under membrane tension}
Before performing a correlation analysis dependent on lag, it is necessary to clarify the definition of “lag time” used in this study. Under membrane tension, the RMSD time series are sampled every 20 ps; therefore, one lag step or $\tau = 1$ corresponds to a delay of 20 ps. The maximum lag $\tau_{\max}$ determines the temporal range over which PCMCI considers historical cross-domain influences during causal inference, that is, the longest lagged dependence allowed in the model.

Figure S3 shows how the pairwise correlations of standardized residuals change with increasing lag under membrane tension. Nearly all domain pairs demonstrate a sharp decline in correlation within the first two lag steps, whereas correlations at larger lags fluctuate around zero. This implies that cross-domain interactions in Piezo1 occur on short time scales, and the system maintains only short-term dynamical memory. Consequently, lagged dependencies beyond two steps do not carry structurally meaningful information. Based on this decay pattern, selecting $\tau_{\max}=2$ provides a reasonable upper bound that captures the relevant temporal dependencies, while larger lag orders might  introduce noise-driven parameters.

To evaluate the sensitivity of the inferred structure to lag time selection, Figure S4 compares PCMCI causal networks obtained using different maximum lag ($\tau_{\max}=1 \sim 6$). Across all parameter settings, the core topology of the network remains highly consistent. The Blade domain is identified as the origin of the network, exerting near-simultaneous causal influences on Beam, Anchor, CTD, OH, and IH. This robustness indicates that Blade is the primary source of cross-domain signal propagation under membrane tension, independent of the choice of lag window. The dominant information pathway described in the manuscript, Blade $\rightarrow$  Beam $\rightarrow$ Anchor $\rightarrow$ CTD $\rightarrow$ IH，is also independent of the choice of  $\tau_{\max}$. Only minor fluctuations in interaction strength are observed, while the direction, sequence, and topology of the pathway remain unchanged. This confirms that the pathway is an intrinsic, essential mechanism of force transmission in Piezo1 under membrane tension. Additional auxiliary pathways (such as Blade $\rightarrow$ Beam $\rightarrow$ Anchor $\rightarrow$ IH and Blade $\rightarrow$ Beam $\rightarrow$ OH $\rightarrow$ IH) are also detected for different lag settings, demonstrating that information propagation in Piezo1 is achieved through multiple parallel routes rather than a single linear chain, thereby enhancing the robustness of the system under membrane tension.

As the maximum lag time increases, the overall topology of the network remains consistent, but the strengths of the causal links gradually weaken. When the lag time exceeds 3 steps, several medium-strength cross-domain interactions are no longer detected because their statistical significance diminishes; only the strongest and most robust connections remain significant. In contrast, $\tau=1$ introduces the least noise but captures only instantaneous effects and may miss short-delay dependencies. A lag time of $\tau=2$ is able to find both instantaneous and short-delay interactions without the weakening effects associated with longer lag windows, thereby providing the best balance between completeness and reliability.

In summary, the results of Figures S3 and S4 show that, under membrane tension, domain-level cascading interactions in Piezo1 occur on a very short temporal scale. The key causal pathways remain highly robust across the entire range of $\tau=1 \sim 6$, while $\tau=2$ provides the most accurate and robust representation of this intrinsic structure. Therefore, $\tau_{\max}=2$ is adopted for PCMCI causal inference under membrane tension. In addition, Table S1 provides the autocorrelation (auto-MCI) and cross-domain causal interaction (cross-MCI) strengths obtained under this lag setting. The numbers in parentheses indicate the actual lag at which each causal link is detected.

\begin{figure}[H]
	\centering
	\includegraphics[scale=0.3]{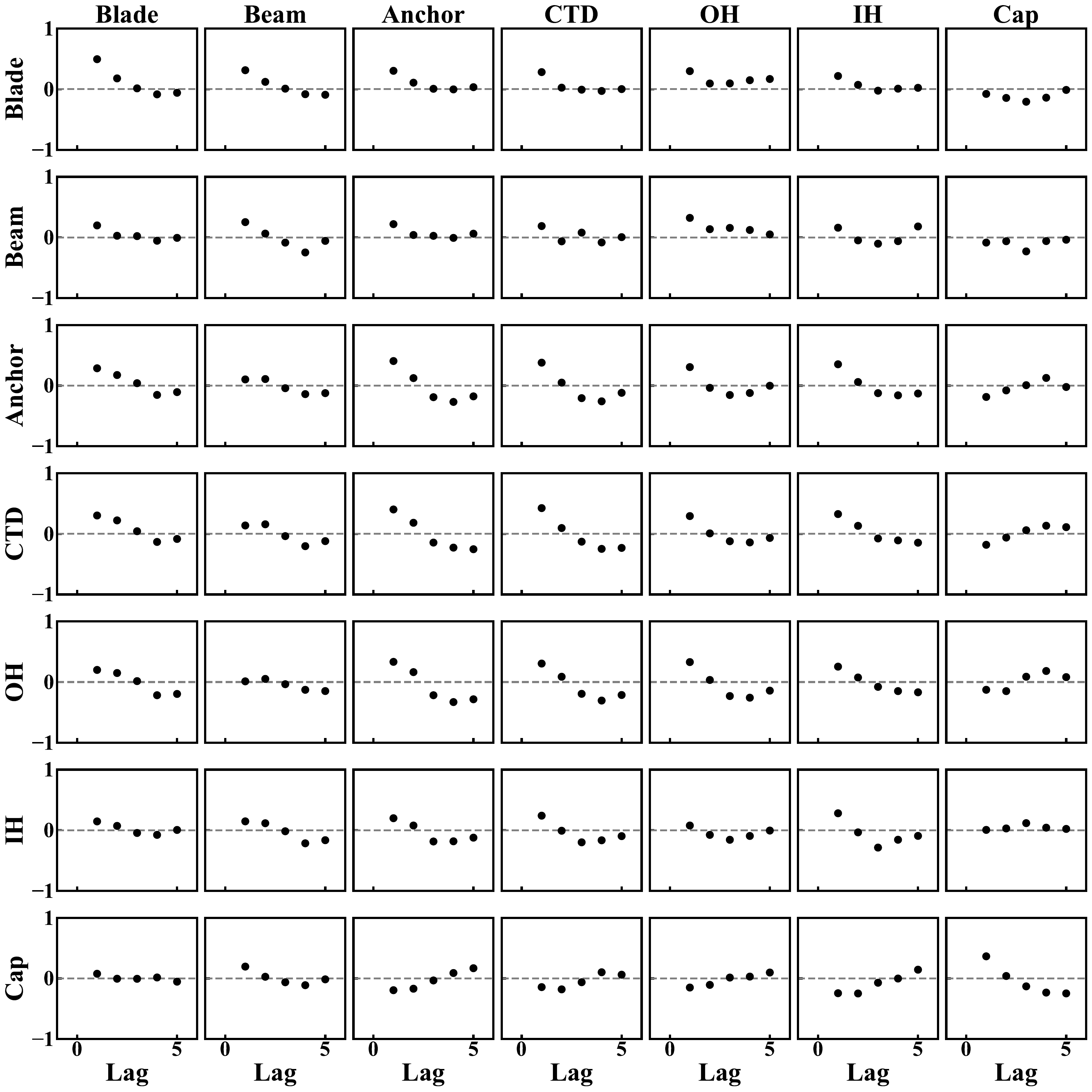}
	\caption*{Figure S3: Lag-dependent pairwise correlations of the standardized residuals under membrane tension. The scatter points depict how correlation strengths change across lag steps and serve as the basis for selecting the maximum lag parameter $\tau_{\max}$ in PCMCI.}
	\label{fg:S3}
\end{figure}

\begin{figure}[H]
	\centering
	\includegraphics[scale=0.2]{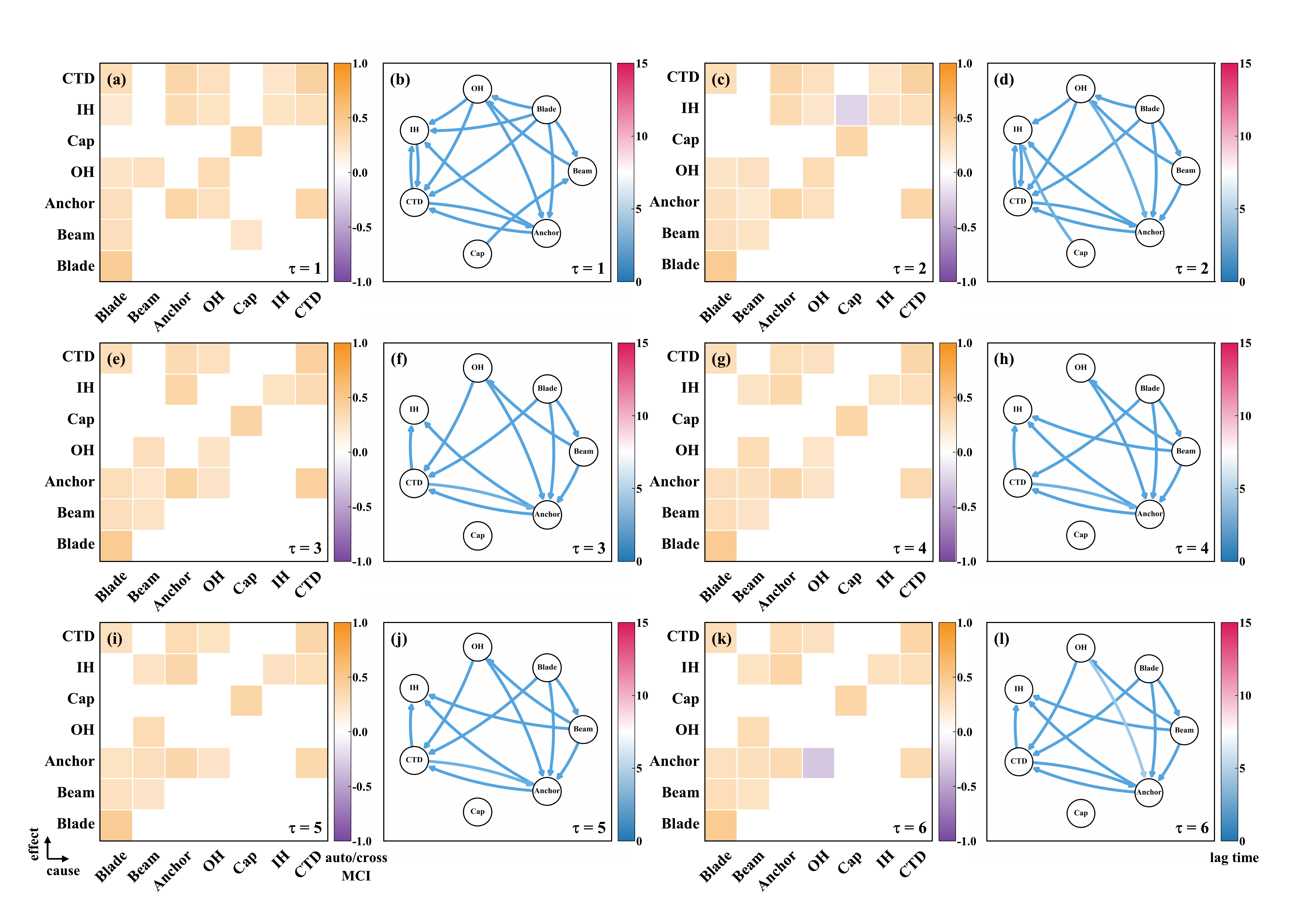}
	\caption*{Figure S4: Allosteric communication network of Piezo1 under membrane tension obtained using different maximum lag times ($\tau_{\max}=1\sim 6$). Panels (a, c, e, g, i, k) show the autocorrelation (auto-MCI) and cross-domain causal interaction (cross-MCI) strengths for each $\tau_{\max}$. Orange indicates positive correlation, purple indicates negative correlation, and the color intensity reflects the interaction strength. Auto-MCI quantifies the influence of a domain’s own history on its current state, whereas cross-MCI captures the influence of other domains’ histories on the current state. Panels (b, d, f, h, j, l) display the corresponding directed causal networks. Arrow directions indicate causal flow (cause → effect), and arrow colors stand for the lag time associated with each causal link.}
	\label{fg:S4}
\end{figure}

\begin{table}[H]
	\centering
        \caption*{Table S1: Causal strengths between Piezo1 domains under membrane tension with $\tau_{\max}=2$. Numbers in parentheses denote the lag at which the corresponding causal link is detected.}
	\vspace{0 em} 
	\setlength{\tabcolsep}{5pt} 
	\begin{tabular}{c|ccccccc}
        \hline
        \diagbox{Cause}{Effect} & Blade & Beam & Anchor & CTD & OH & IH & Cap \\
        \hline
        Blade & 0.467 (1) & 0.293 (1) & 0.289 (1) & 0.311 (1) & 0.241 (1) & -- & -- \\
        \hline
        Beam & --  & 0.259 (1) & 0.215 (1) & -- & 0.271 (1) & -- & -- \\
        \hline
        Anchor & -- & -- & 0.386 (1) & 0.370 (1) & -- & 0.333 (1) & -- \\
        \hline
        CTD & -- & -- & 0.378 (1) & 0.411 (1) & -- & 0.307 (1) & -- \\
        \hline
        OH & -- & -- & 0.248 (2) & 0.283 (1) & 0.321 (1) & 0.227 (1) & -- \\
        \hline
        IH & -- & -- & -- & 0.234 (1) & -- & 0.271 (1) & -- \\
        \hline
        Cap & -- & -- & -- & -- & -- & -0.232 (1) & 0.376 (1) \\
        \hline
\end{tabular}
\end{table}

\section*{C. Conformational responses of the membrane protein under shockwave impact with and without cavitation}
Under shockwave impact, we compared the effects of pure shock impact (without cavitation) and shock-induced cavitation (bubble radius $R = 10\text{nm}$) on membrane deformation and Piezo1 conformational responses. Figure S5(a) shows the minimum membrane thickness ($h_{\min}$) and the maximum RMSD as functions of particle velocity $u_p$. In both cavitation and non-cavitation scenarios, membrane thickness decreases with increasing $u_p$, indicating that shock impact is able to induce membrane compression. However, the reduction is much more significant in the cavitation case, because of the additional transient high pressure generated during bubble collapse. This enhanced local compression will disturb the membrane-protein interface and create the mechanical conditions necessary for subsequent conformational rearrangement. Meanwhile, the behaviors of the maximum RMSD for the two cases are quite different. Without cavitation,  the maximum RMSD of Piezo1 are around 1.0~nm and remains structurally stable. In contrast, RMSD rises rapidly with increasing $u_p$ under cavitation, signifying the substantial conformational reorganization which may involve coordinated motions of multiple transmembrane helices.

Figure~S5(b) presents the changes in the projected area (defined in Figure 1b) of the three-blade region and the pore radius. In the non-cavitation case, the projected area remains nearly unchanged across the simulated $u_p$ range, and the pore radius stays at $2\sim 3~nm$, suggesting that the three-blade structure preserves its closed configuration. Under cavitation, however, the projected area expands significantly with increasing $u_p$, shown by the outward splaying of the three blades. The pore radius also increases profoundly and reaches approximately $5~nm$ at $u_p = 0.8 \text{km/s}$, as reported in the literature for activating Piezo1 \cite{WOS:000613509200008}. At higher shock intensities, the pore continues to enlarge, indicating that cavitation not only triggers gating but also modulates the extent of channel opening.

Shock-induced cavitation plays a crucial role in the membrane protein’s response. Compared with pure shock impact, cavitation will enhance membrane thinning and global protein perturbation, and significantly promotes blade expansion and pore opening. This constitutes the key physical mechanism by which Piezo1 transitions from the closed state to the activated state.

\begin{figure}[H]
	\centering
	\includegraphics[scale=0.7]{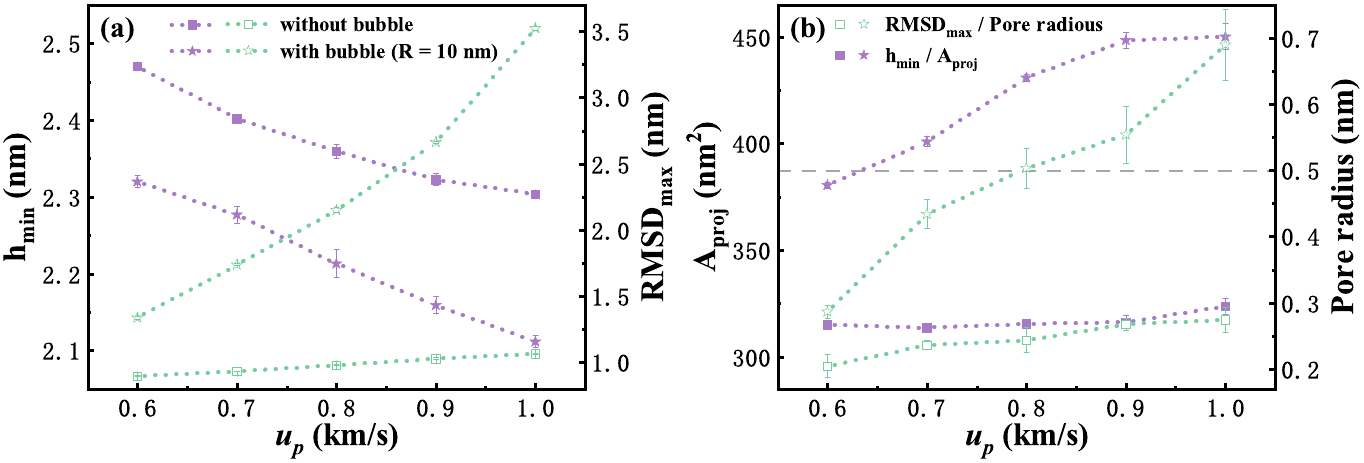}
	\caption*{Figure S5: (a) Variations of the minimum membrane thickness ($h_{\min}$) and the maximum RMSD as functions of the particle velocity $u_p$. (b) Variations of the projected area ($A_{\mathrm{proj}}$) and pore radius with increasing particle velocity $u_p$. The gray line denotes the gating threshold: values above this line indicate that the channel opens, whereas values below it correspond to the non-activated state.}
	\label{fg:S5}
\end{figure}

\section*{D. Validation of shockwave modeling}
The relationship between particle velocity and shock velocity provides an essential validation for shockwave simulations. As shown in Figure S6, the shock velocities obtained in this study is consistent with previous results from experiments \cite{rybakov1996phase}, all-atom molecular dynamics \cite{choubey2011poration, vedadi2010structure}, and MARTINI coarse-grained simulations\cite{santo2014shock}. It can be found that, for the simulations with different spatial resolutions and force-field descriptions, it yields a nearly linear dependence of $u_s$ on $u_p$. The ability of our model to reproduce this relation demonstrates that the generation and propagation of the shock front are physically validated in our simulation setup.

For the simulated range in this study, the smallest deviation from previous results occurs at a particle velocity of $u_p = 1~\mathrm{km/s}$, i.e., our system’s response is well consistent with reported shock behavior. Therefore, $u_p = 1~\mathrm{km/s}$ is selected as the representative impact condition for the subsequent structural and causal analyses of Piezo1 in the present study.

\begin{figure}[H]
	\centering
	\includegraphics[scale=0.4]{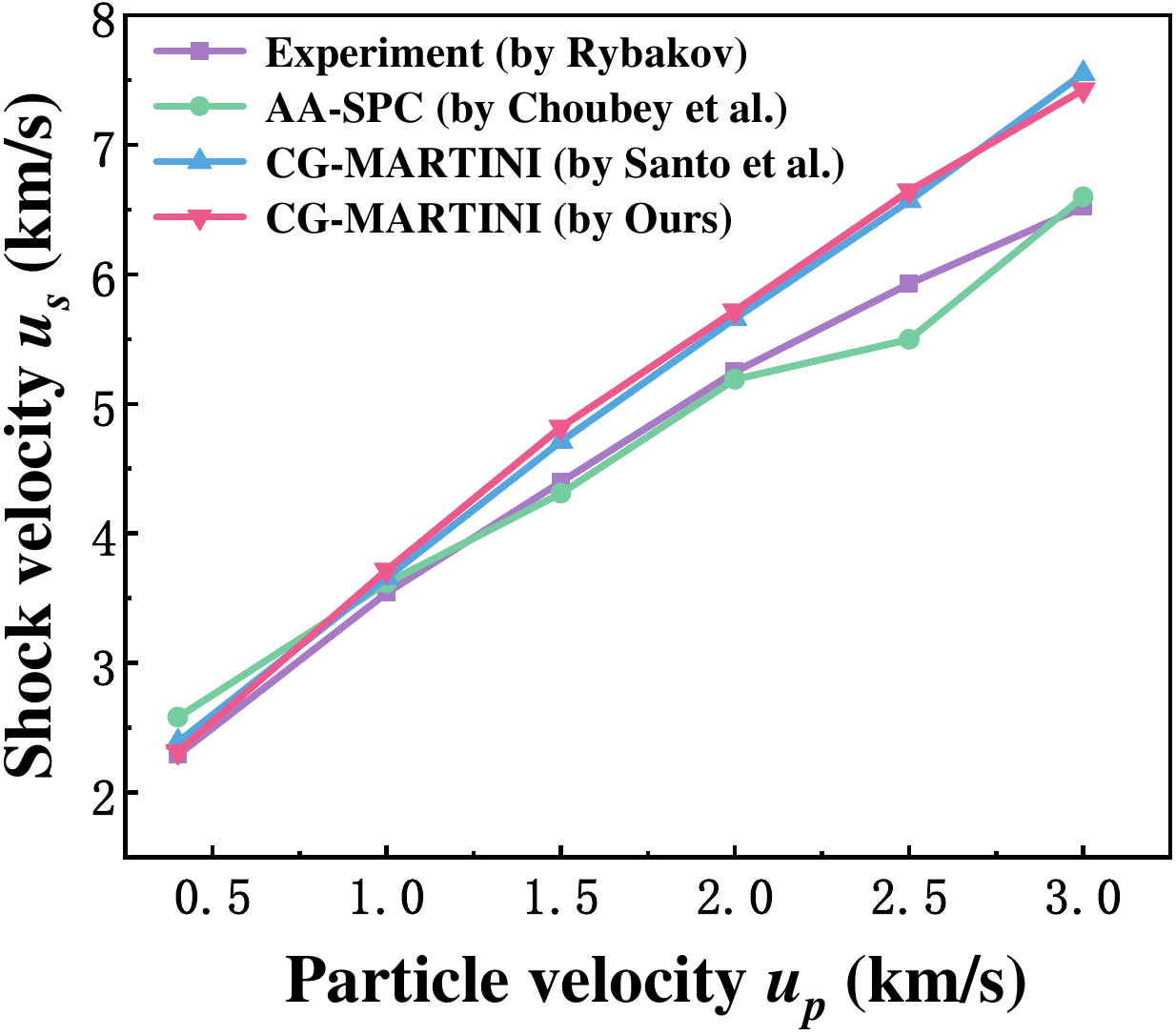}
	\caption*{Figure S6: Shock velocity as a function of particle velocity $u_p$, compared with previous experiment (Rybakov \cite{rybakov1996phase}), all-atom simulation (Choubey et al. \cite{choubey2011poration}), and MARTINI coarse-grained results (Santo et al. \cite{santo2014shock}).}
	\label{fg:S6}
\end{figure}

\section*{E. Identification of the compression–tension transition under shockwave impacting}
The trend components extracted from the STL decomposition, together with their first derivatives, are used to locate the turning point of the RMSD evolution under shockwave impact (Figure S7). It can be found that the responses among domains are different. Blade, Beam, and Cap show a monotonic increase throughout the loading process without a obvious extremum near 25 ps. In comparison, Anchor and CTD begin to decrease around 25 ps with zero first-derivative, while OH and IH display smaller but still recognizable turning features at approximately $t=25$ ps.

Therefore, these observations show that the system as a whole switches from a compression-dominated regime to a tension-driven re-expansion around 25 ps. Although some peripheral domains (Blade, Beam, Cap) continue to increase monotonically, their behavior does not contradict the global transition, which is strongly supported by the turning points in Anchor and CTD and, albeit weaker, features in OH and IH. Therefore, it is reasonable to identify 25 ps as a physically meaningful compression–tension transition point.

\begin{figure}[H]
	\centering
	\includegraphics[scale=0.8]{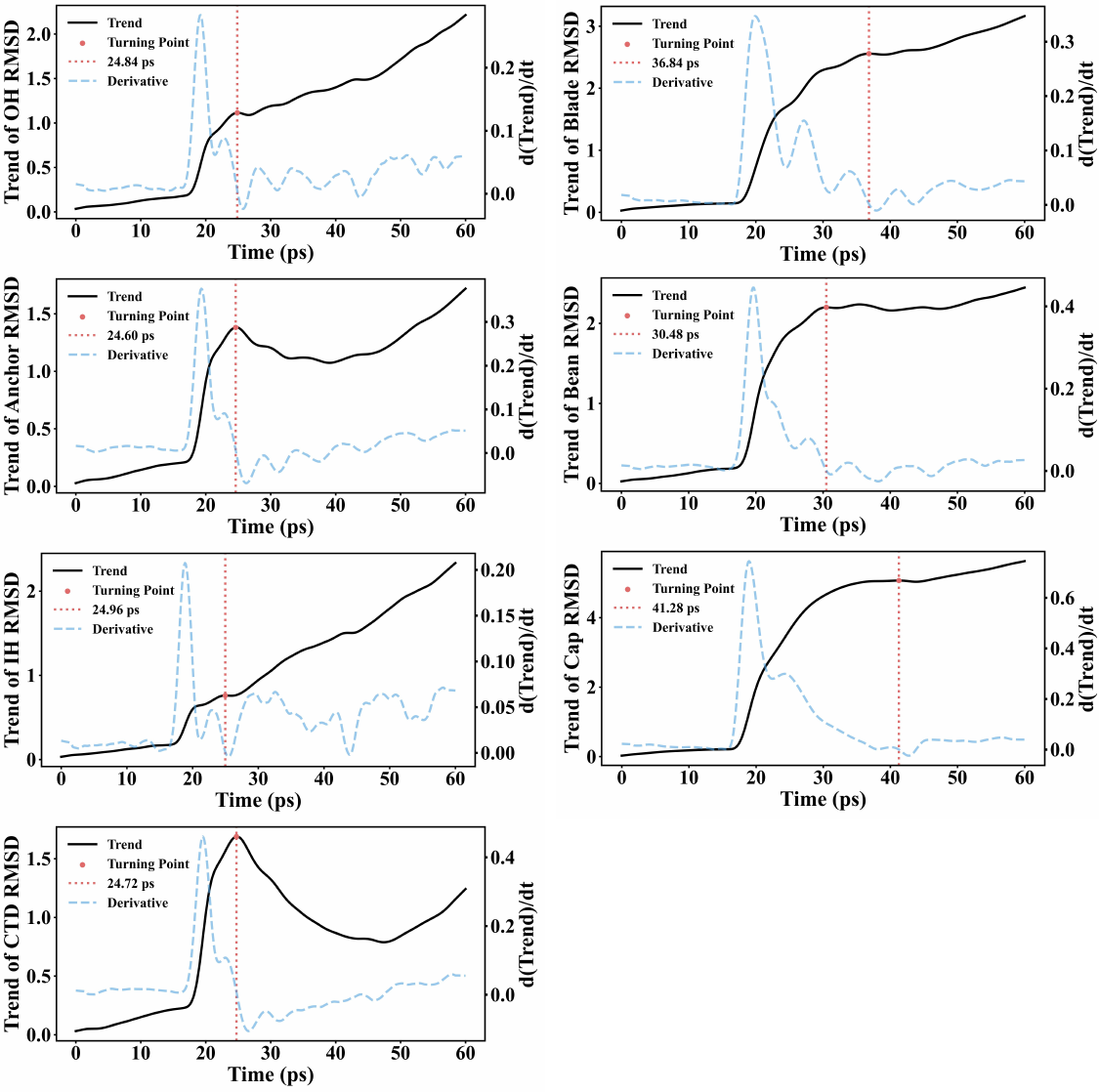}
	\caption*{Figure S7: Trend components of the RMSD time series and their first derivatives obtained from STL decomposition, used to identify the compression–tension transition point under shockwave impacting.}
	\label{fg:S7}
\end{figure}

\section*{F. Selection of $\tau_{\max}$ and robustness of PCMCI causal networks during shockwave compression}
Shockwave simulations are performed with an RMSD sampling interval of 0.12 ps, so one lag step corresponds to a 0.12 ps time delay. Figure S8 shows the pairwise correlations of the standardized residuals for different lag steps. Unlike the membrane-tension case, where the correlations for all the domains decay rapidly, the dependence of the correlation on the lag time is quite different among the domain pairs. For some domain pairs, the correlation weaken quickly after $1\sim 3$ lag steps, whereas others maintain strong correlations even at larger lags. This phenomenon manifests the complex temporal characteristics of shock loading, including rapid front propagation, transient compression, and early local rebound. 

Based on this observation, the PCMCI causal networks obtained under different maximum lag values ($\tau_{\max}=1\sim 6$) are analyzed in Figure S9 to determine the effective temporal window of cross-domain coupling. When $\tau_{\max}=1$, the network shows a diverse array of cross-domain causal connections. As $\tau_{\max}$ increases, many moderate-strength links observable at $\tau_{\max}=1$ disappear, thus the networks become increasingly sparse. Importantly, the causal edges preserved at larger $\tau_{\max}$ do not introduce any new information pathways, all of them have been already embedded in the $\tau_{\max}=1$ network. This indicates that the essential causal structure in the compression stage is established within a single lag step (0.12 ps). Longer temporal memory does not contribute additional structural information, instead, it leads to the loss of causal edges due to reduced statistical significance.

It can be found that, for different $\tau_{\max}$, Blade always emerges as the initiator of conformational change, transmitting signals to Beam, Anchor, CTD, OH, and IH. The lever-like pathway, Blade → Beam → Anchor → CTD → OH/IH, is present in the $\tau_{\max}=1$ network. This demonstrates that even under transient shock compression, the cross-domain communication architecture of Piezo1 same fundamental logic as under membrane tension is still on action. In addition, a rapid allosteric pathway originating from Cap also appears recurrently across multiple $\tau_{\max}$ values, suggesting that under strong transient compression, the Cap domain may participate in earlier transmitting conformational perturbations and provide an additional fast-acting trigger toward channel activation.

Therefore, Figures S8 and S9 show that cross-domain coupling during the shock-compression stage of Piezo1 is established within an short timescale defined by $\tau_{\max}=1$ (0.12 ps). The dominant causal pathways remain topologically robust across $\tau_{\max}=1\sim 6$, while $\tau_{\max}=1$ preserves all essential connections and avoids noise-induced edge loss at larger lag windows. Therefore, $\tau_{\max}=1$ is adopted for PCMCI causal inference during the shockwave compression stage, with the corresponding autocorrelation (auto-MCI) and cross-domain causal interaction (cross-MCI) strengths listed in Table S2.

\begin{figure}[H]
	\centering
	\includegraphics[scale=0.3]{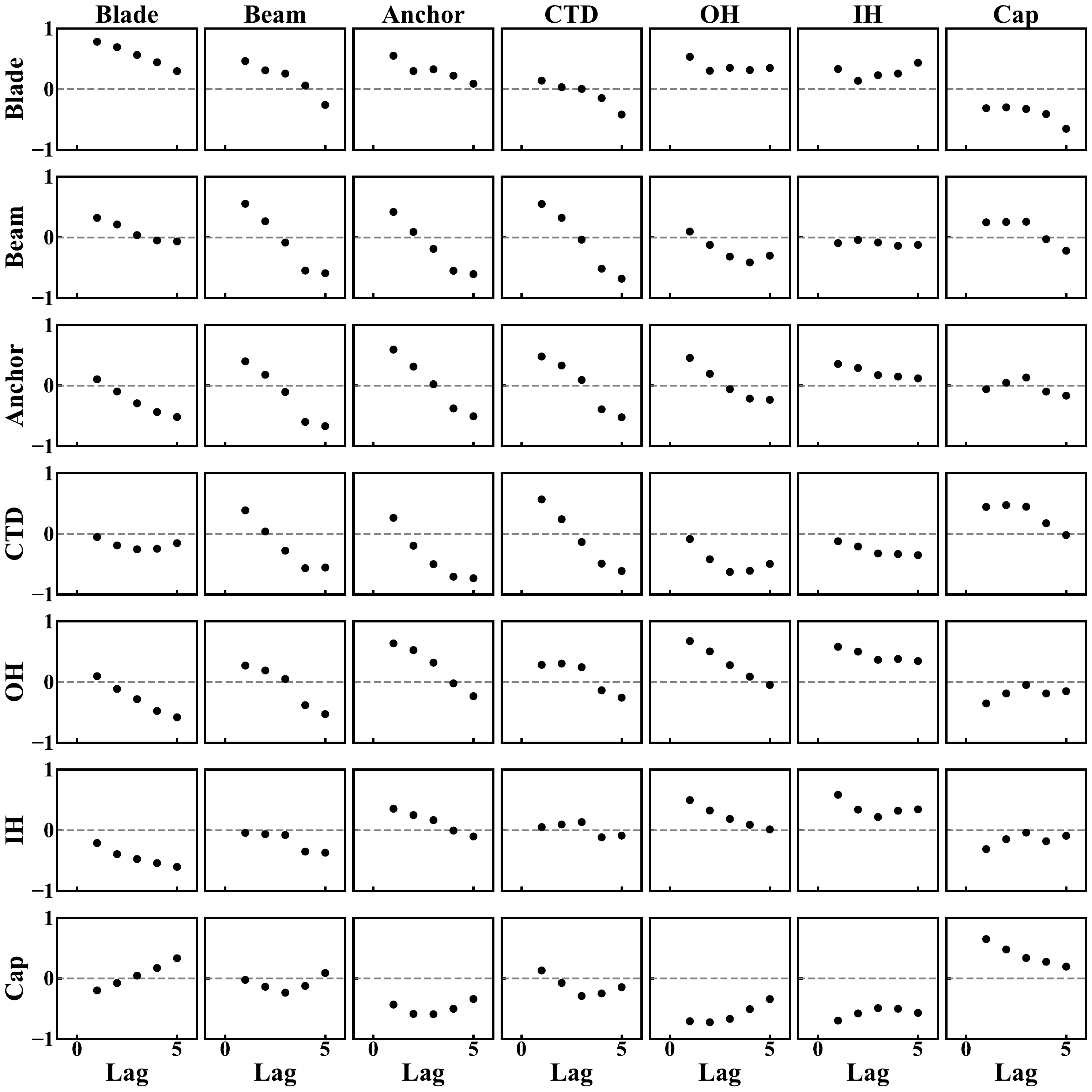}
	\caption*{Figure S8: Lag-dependent pairwise correlations of the standardized residuals under shockwave compression. The scatter points depict how correlation strengths change across lag steps and serve as the basis for selecting the maximum lag parameter $\tau_{\max}$ in PCMCI.}
	\label{fg:S8}
\end{figure}

\begin{figure}[H]
	\centering
	\includegraphics[scale=0.2]{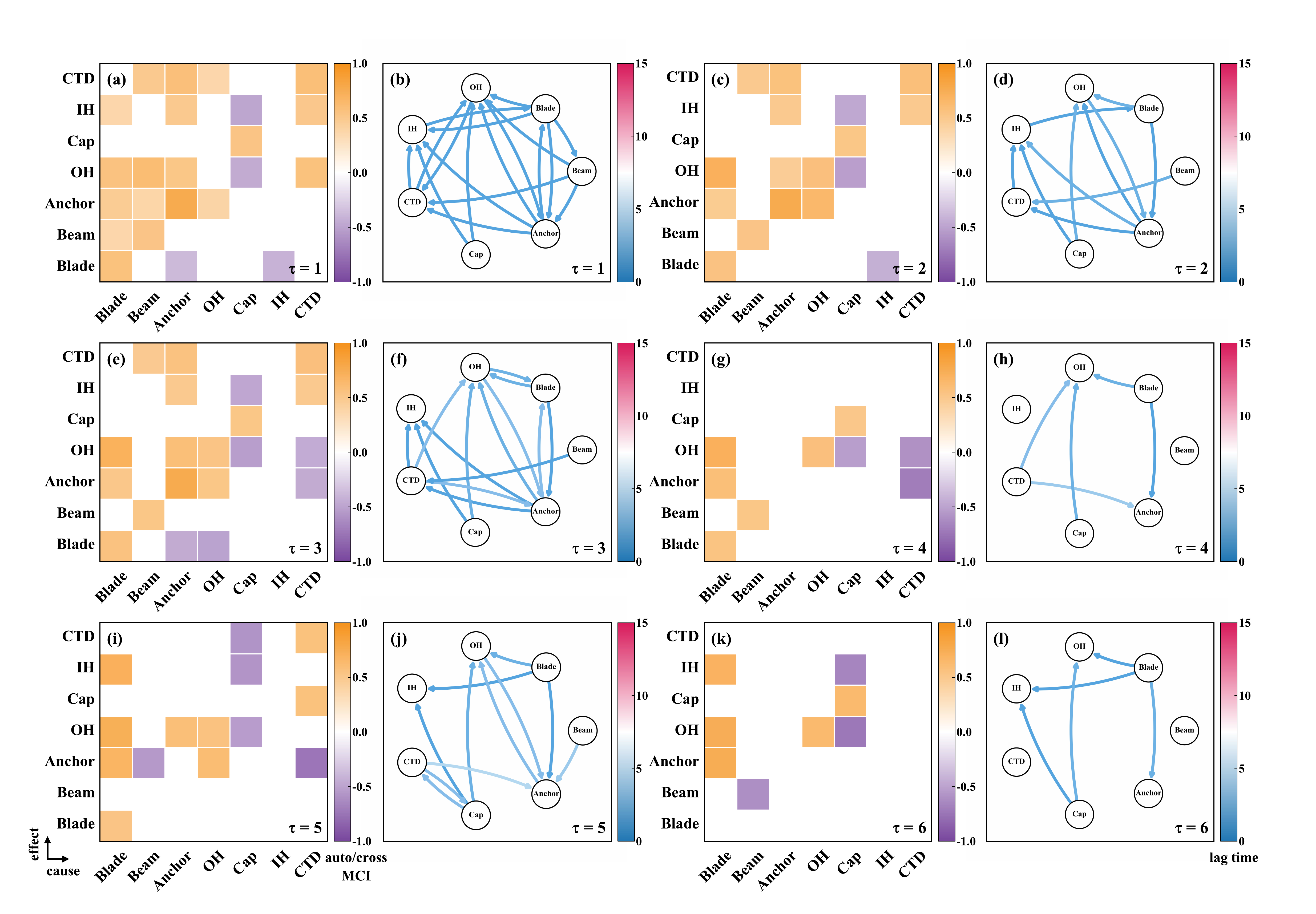}
	\caption*{Figure S9: Allosteric communication network of Piezo1 under shockwave compression obtained using different maximum lag times ($\tau_{\max}=1\sim 6$). Panels (a, c, e, g, i, k) show the autocorrelation (auto-MCI) and cross-domain causal interaction (cross-MCI) strengths for each $\tau_{\max}$. Orange indicates positive correlation, purple indicates negative correlation, and the color intensity reflects the interaction strength. Panels (b, d, f, h, j, l) display the corresponding directed causal networks. Arrow directions indicate causal flow (cause → effect), and arrow colors stand for the lag time associated with each causal link.}
	\label{fg:S9}
\end{figure}

\begin{table}[H]
	\centering
        \caption*{Table S2: Causal strengths between Piezo1 domains under shockwave compression with $\tau_{\max}=1$. Numbers in parentheses denote the lag at which the corresponding causal link is detected.}
	\vspace{0 em} 
	\setlength{\tabcolsep}{5pt} 
	\begin{tabular}{c|ccccccc}
	\hline
	\diagbox{Cause}{Effect} & Blade & Beam & Anchor & CTD & OH & IH & Cap \\
	\hline
	Blade & 0.565 (1) & 0.369 (1) & 0.464 (1) & -- & 0.559 (1) & 0.372 (1) & -- \\
	\hline
	Beam & --  & 0.532 (1) & 0.382 (1) & 0.494 (1) & 0.617 (1) & -- & -- \\
	\hline
	Anchor & -0.376 (1) & -- & 0.766 (1) & 0.580 (1) & 0.523 (1) & 0.489 (1) & -- \\
	\hline
	CTD & -- & -- & -- & 0.586 (1) & 0.564 (1) & 0.502 (1) & -- \\
	\hline
	OH & -- & -- & 0.393 (1) & 0.374 (1) & -- & -- & -- \\
	\hline
	IH & -0.417 (1) & -- & -- & -- & -- & -- & -- \\
	\hline
	Cap & -- & -- & -- & -- & -0.457 (1) & -0.486 (1) & 0.537 (1) \\
	\hline
\end{tabular}
\label{tb:S2}
\end{table}

\section*{G. Selection of $\tau_{\max}$ and robustness of PCMCI causal networks during shockwave tension}
As we have seen, under the membrane-tension case, the correlations between domains of Piezo1 decay rapidly with increasing delay time. In the shock-compression stage of shock impacting, its essential structure of allosteric network are established within an short temporal window (approximately one lag or 0.12ps). Different from these phenomenon, substantially longer delay-dependent correlations are found in the tension stage under shock impacting, as shown in Figure S10. A large subset of domain pairs have non-zero correlation coefficients for 10–15 delay steps (approximately 1.2–1.8 ps), indicating that cross-domain coupling during the tension stage involves a considerably longer temporal correlation scale. Although the correlation persistence illustrated in Figure S10 provides a wide range as candidate for selecting the delay-time window, the definitive choice of delay time cannot rely on correlation analysis alone. Therefore, we examine the PCMCI causal networks obtained under different maximum delay times ($\tau_{\max}=13\sim 18$) in Figure S11 to evaluate the robustness of cross-domain causal structure within this candidate delay range.

Our results show that, in the range of $13\sim 18$ delay steps, the essential causal pathways remain highly consistent in network topology, implying that the causal structure during the tension stage is robust within this delay interval. However, when the maximum delay time is too short (e.g., $\tau_{\max}=13$–14), some cross-domain interactions with longer delays in Figure S10 cannot be captured. Conversely, further increasing the maximum delay time (e.g., $\tau_{\max}=17$–18) leads to the disappearance of the medium-strength cross-domain interactions. Therefore, by considering the delay-dependent correlation range from Figure S10 and the causal-network robustness depicted in Figure S11, it is found that $\tau_{\max}=15$ provides the optimal balance balance the need of capturing meaningful delayed interactions while avoiding unnecessarily long delays that do not contribute valid causal structure, and is thus selected as the maximum delay time for PCMCI causal inference during the shockwave tension stage, with the corresponding autocorrelation (auto-MCI) and cross-domain causal interaction (cross-MCI) strengths listed in Table S3.

\begin{figure}[H]
	\centering
	\includegraphics[scale=0.3]{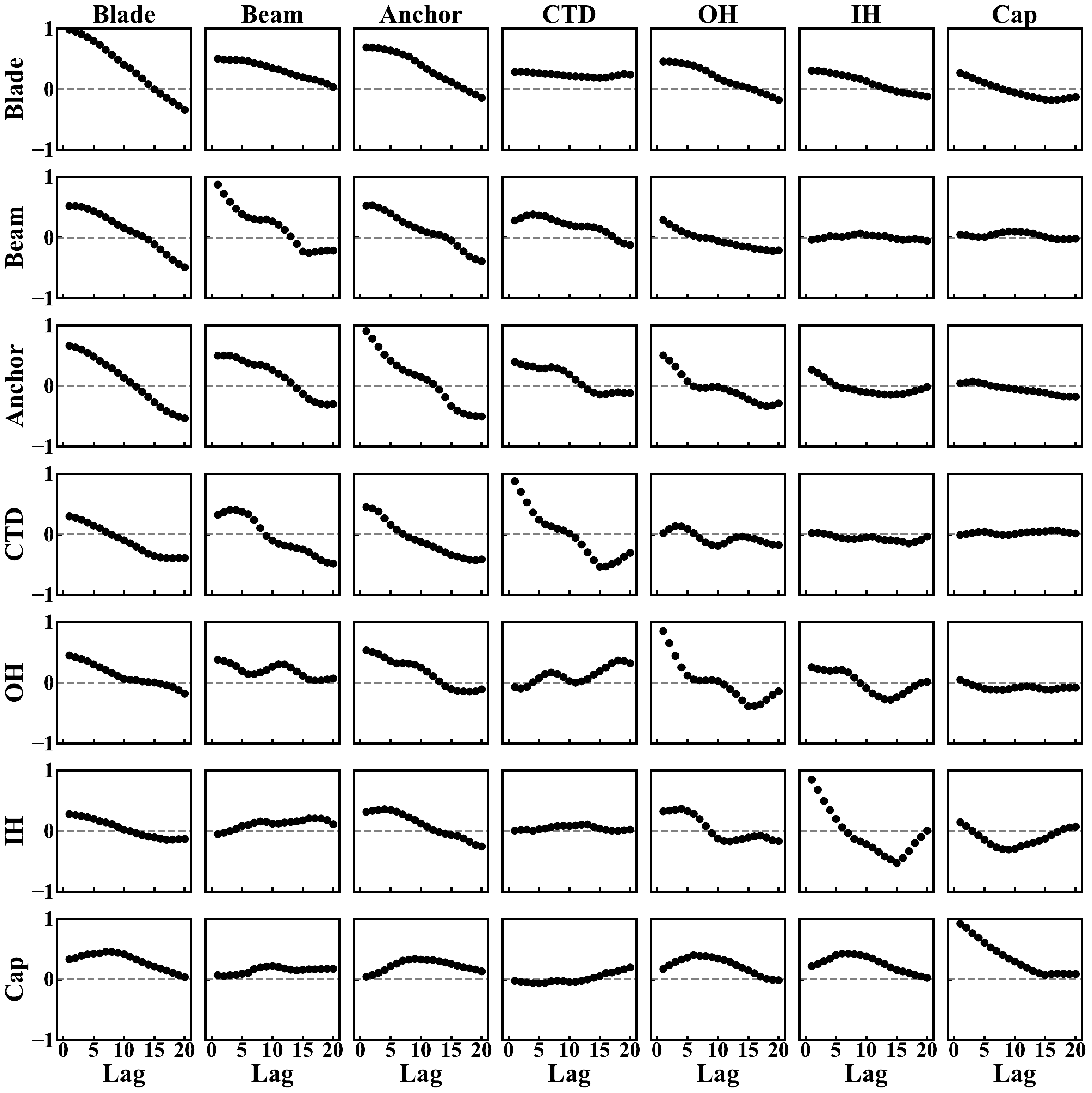}
	\caption*{Figure S10: Lag-dependent pairwise correlations of the standardized residuals under shockwave tension. The scatter points depict how correlation strengths change across lag steps and serve as the basis for selecting the maximum lag parameter $\tau_{\max}$ in PCMCI.}
	\label{fg:S10}
\end{figure}

\begin{figure}[H]
	\centering
	\includegraphics[scale=0.2]{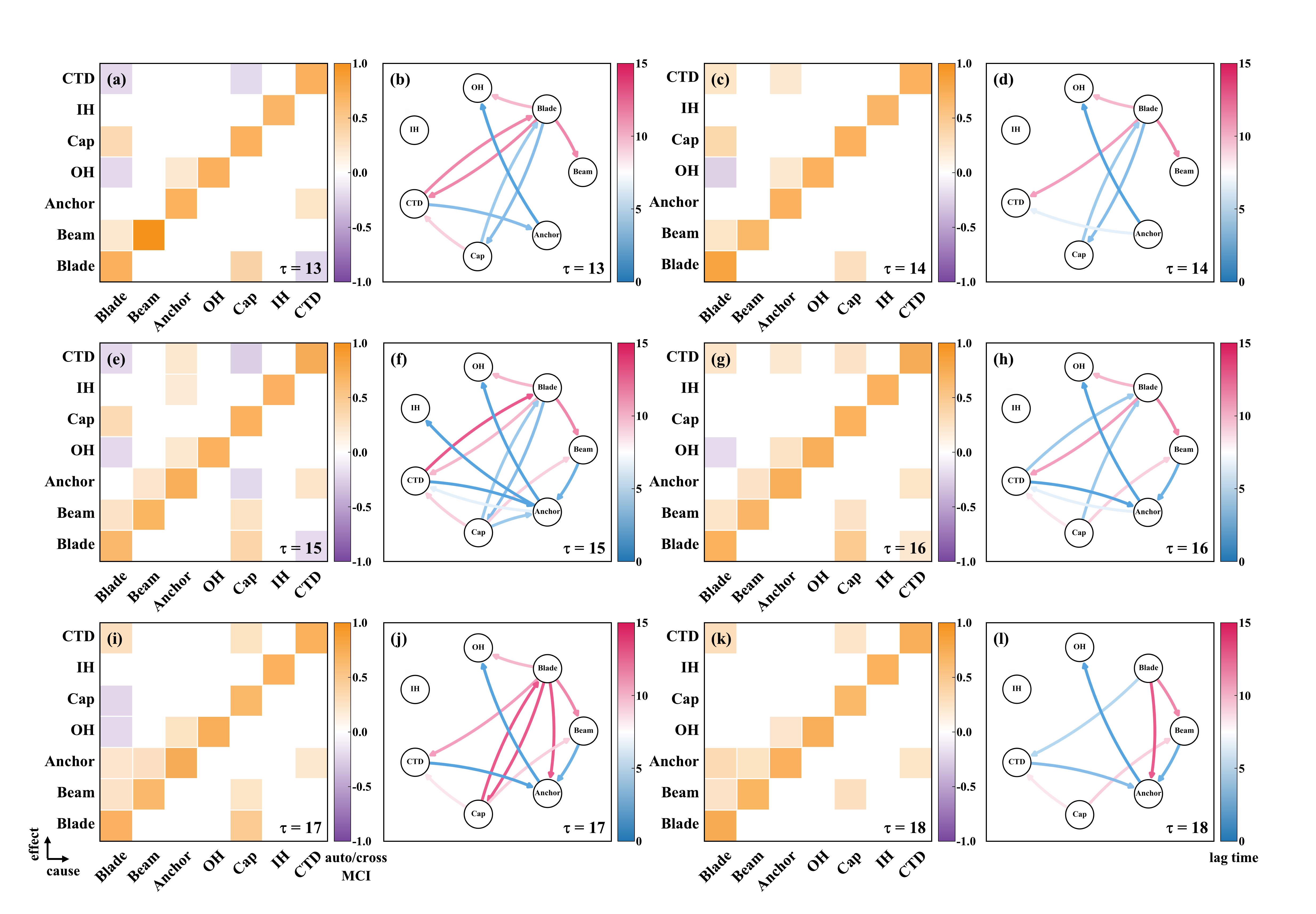}
	\caption*{Figure S11: Allosteric communication network of Piezo1 under shockwave tension obtained using different maximum lag times ($\tau_{\max}=13\sim 18$). Panels (a, c, e, g, i, k) show the autocorrelation (auto-MCI) and cross-domain causal interaction (cross-MCI) strengths for each $\tau_{\max}$. Orange indicates positive correlation, purple indicates negative correlation, and the color intensity reflects the interaction strength. Panels (b, d, f, h, j, l) display the corresponding directed causal networks. Arrow directions indicate causal flow (cause → effect), and arrow colors stand for the lag time associated with each causal link.}
	\label{fg:S11}
\end{figure}

\begin{table}[H]
	\centering
        \caption*{Table S3: Causal strengths between Piezo1 domains under shockwave tension with $\tau_{\max}=15$. Numbers in parentheses denote the lag at which the corresponding causal link is detected.}
	\vspace{0 em} 
	\setlength{\tabcolsep}{3pt} 
	\begin{tabular}{cccccccc}
		\hline
		\diagbox{Cause}{Effect} & Blade & Beam & Anchor & CTD & OH & IH & Cap \\
		\hline
		Blade & 0.655 (1) & 0.252 (13) & -- & -0.202 (11) & -0.208 (11) & -- &  \begin{tabular}{@{}c@{}}0.326 (3) \\ 0.200 (1) \end{tabular} \\
		\hline
		Beam & --  & 0.687 (1) & 0.227 (2) & -- & -- & -- & -- \\
		\hline
		Anchor & -- & -- & 0.723 (1) & 0.211 (7) & 0.210 (1) & 0.191 (1) & -- \\
		\hline
		CTD & -0.192 (15) & -- & 0.232 (1) & 0.740 (1) & -- & -- & -- \\
		\hline
		OH & -- & -- & -- & -- & 0.704 (1) & -- & -- \\
		\hline
		IH & -- & -- & -- & -- & -- & 0.693 (1) & -- \\
		\hline
		Cap & 0.383 (4) & 0.256 (10) & -0.198 (4) & -0.260 (10) & -- & -- & 0.703 (1) \\
		\hline
	\end{tabular}
	\label{tb:S3}
\end{table}

\bibliography{SI}